\documentclass[12pt]{article}
\usepackage[pdftex]{pict2e}

\usepackage{mathrsfs,amsthm,amsmath,latexsym,amssymb,amsfonts,amscd,graphicx,latexsym}
\usepackage{graphics,lscape,fancyhdr,array,stmaryrd,euscript}
\usepackage{ifthen,empheq,slashed}
\usepackage{color}
\numberwithin{equation}{section}
\usepackage{setspace}
\usepackage{cite}

\DeclareFontFamily{U}{mathx}{\hyphenchar\font45 }
\DeclareFontShape{U}{mathx}{m}{n}{
	<-> mathx5
}{}
\DeclareSymbolFont{mathx}{U}{mathx}{m}{n}
\DeclareMathAccent{\wideTilde}{0}{mathx}{"72}

\usepackage{bm} 

\usepackage[hang,flushmargin]{footmisc} 
\usepackage[citecolor=purple,linkcolor=blue,urlcolor=blue,colorlinks=true,filecolor=green,
pdfstartview=FitV, 
bookmarksopen=true,
hyperfootnotes=true]{hyperref}

\usepackage{makecell}
\usepackage{geometry}

\newgeometry{left=.8in,right=.8in,top=.8in,bottom=.85in}
\usepackage[most]{tcolorbox}
\usepackage{enumitem}
\usepackage{footnotebackref}

\usepackage{soul}

\usepackage{framed}

\usepackage{tikz}
\usepackage{xcolor}
\usepackage{colortbl}
\usepackage{pgfplots}
\pgfplotsset{compat=1.16}

\definecolor{galilei}{RGB}{255,255,153}
\definecolor{minkowski}{RGB}{204,255,204}
\definecolor{carroll}{RGB}{255,204,204}

\usetikzlibrary{calc}
\usetikzlibrary{arrows.meta, positioning}

\newcommand{\xdownarrow}[1]{%
	{\left\Downarrow\vbox to #1{}\right.\kern-\nulldelimiterspace}
}

\usepackage{multirow}
\usepackage{array}
\usepackage{makecell}

\definecolor{yell}{cmyk}{0,0,0,1}

\usepackage{lipsum} 
\usepackage{abstract}

\usepackage[retainorgcmds]{IEEEtrantools}
\def\beak{\begin{IEEEeqnarray*}}
	\def\eeak{\end{IEEEeqnarray*}}
\def\nk{\IEEEyesnumber\phantomsection}

\newcommand{\bes}[1]{\begin{subequations}\label{#1} }
	\newcommand{\ees}{\end{subequations}}

\def\be {\begin{equation}}
	\def\ee {\end{equation}}
\def\le {\left}
\def\ri {\right}
\def\p {\partial}
\def\d {\delta}

\def\g {\gamma}
\def\c {_\mathrm{c}}
\def\pc {_\mathrm{pc}}

\def\1{_{_1}}
\def\2{_{_2}}

\begin{document}


\begin{center}
	
	{\bfseries \large{Post-Carroll Algebra, Conformal Extensions, and Field Theories}} \vspace{-8pt}
	\par\noindent\rule{465pt}{2pt}\\
	
	\vskip 0.04\textheight
	
	
		Mojtaba \textsc{Najafizadeh}{}

	\vskip 0.01\textheight

		\vspace*{5pt}
		{\small\em 
				School of Physics, Institute for Research in Fundamental Sciences (IPM), \\ 
				P.O.Box 19395-5531, Tehran, Iran}
	
	\vskip 0.01\textheight

	\vskip 0.01\textheight
	
	\href{mailto:mnajafizadeh@ipm.ir}{mnajafizadeh@ipm.ir}
	
	\vskip 0.05\textheight

{\bf Abstract:}\end{center}
\vskip -0.01\textheight
By incorporating leading $c\,$-dependent corrections to the Carroll transformations, we introduce the ``post-Carroll transformations''. We demonstrate that these transformations are consistent with post-Carrollian mechanics \cite{Najafizadeh:2025ksm}; furthermore, they give rise to the so-called ``post-Carroll algebra''. We show that, unlike the Carroll algebra, this new structure allows for a central charge in higher dimensions; we refer to it as the ``Carroll–Bargmann algebra''. To construct conformal extensions, we first build the conformal extension of the post-Carroll algebra and study field theories invariant under this symmetry. We then construct the conformal extension of the Carroll–Bargmann algebra, referred to as the ``Carroll–Schr\"odinger algebra'', and demonstrate that it precisely matches the symmetry algebra of the higher-dimensional Carroll–Schr\"odinger theory \cite{Najafizadeh:2024imn}. Finally, we derive the general form of two-point functions in a post-Carrollian CFT, which in $1+1$ dimensions exhibits both electric and magnetic sectors, while in higher dimensions only the magnetic sector survives.

\vskip 0.03\textheight

\noindent\textsc{Keywords}: {\small Post-Carroll transformations, post-Carroll algebra, Carroll–Bargmann algebra, conformal post-Carroll algebra, Carroll–Schr\"odinger algebra, Carroll–Schr\"odinger equation, correlation functions.}

\setlength{\leftskip}{0cm} 
\setlength{\rightskip}{0cm} 


\vskip 0.03\textheight

\newpage
\small\tableofcontents
\newpage

\section{Introduction}

The Carrollian symmetry is obtained as a contraction of the Poincaré symmetry by sending the speed of light to zero ($c \to 0$). It was first introduced in \cite{Levy1965,SenGupta:1966qer} and has gained increased attention over the past decade following the discovery that the Carrollian conformal algebra is isomorphic to the Bondi–Metzner–Sachs (BMS) algebra \cite{Bondi:1962px,Sachs:1962zza,Sachs:1962wk} in one higher dimension \cite{Duval:2014uoa,Duval:2014uva,Duval:2014lpa}. Carrollian structures have been studied in different contexts and from various aspects, for example, in flat space holography \cite{Barnich:2006av,Bagchi:2016bcd,Ciambelli:2018wre,Gupta:2020dtl,Donnay:2022aba,Bagchi:2022emh,Bagchi:2023cen,Poulias:2025eck,Fiorucci:2025twa}, correlation functions \cite{Nguyen:2023miw,Alday:2024yyj,Afshar:2024llh,Nguyen:2025sqk,Despontin:2025dog,Kulkarni:2025qcx,Marotta:2025qjh}, fractons \cite{Bidussi:2021nmp,Marsot:2022imf,Figueroa-OFarrill:2023vbj,Figueroa-OFarrill:2023qty,Pena-Benitez:2023aat,Ahmadi-Jahmani:2025iqc}, cosmology \cite{deBoer:2021jej,deBoer:2023fnj}, gravity \cite{Henneaux:1979vn,Hartong:2015xda,Bergshoeff:2017btm,Hansen:2021fxi,Guerrieri:2021cdz,Campoleoni:2022ebj,Figueroa-OFarrill:2022mcy,Sengupta:2022rbd,March:2024zck,Ecker:2024czh,Bergshoeff:2024ilz,Concha:2025vhd,Afshar:2025imp}, scalar fields \cite{Henneaux:2021yzg,deBoer:2021jej,Bergshoeff:2022qkx,Rivera-Betancour:2022lkc,Baiguera:2022lsw,Bekaert:2024itn,Sharma:2025rug,Majumdar:2025juk,Bruce:2026knl}, fermions \cite{Banerjee:2022ocj,Bagchi:2022eui,Koutrolikos:2023evq,Bergshoeff:2023vfd,Ekiz:2025hdn,Bagchi:2026lgk,Bagchi:2026emg,Majumdar:2026iol}, and supersymmetry \cite{Bergshoeff:2015wma,Bagchi:2022owq,Koutrolikos:2023evq,Kasikci:2023zdn,Zheng:2025cuw,Bruce:2026yvw,Bulunur:2026yav}. For recent reviews, see e.g. \cite{Bagchi:2025vri,Ciambelli:2025unn,Ruzziconi:2026bix}, and also the recent theses \cite{Ecker:2025vnl,Shekar:2026btr}, and references therein.

The Galilean symmetry, on the other hand, arises as another contraction of the Poincar\'e symmetry, corresponding to sending the speed of light to infinity ($c\to \infty$). It is well known that the Galilei algebra admits a non-trivial central charge $M$ in any spacetime dimension, giving rise to the Bargmann algebra. In contrast, the Carroll algebra admits such a non-trivial central charge only in 1+1 dimensions (see, e.g., \cite{Najafizadeh:2024imn,Ecker:2025ncp}) and lacks it in higher dimensions. This raises the question: how can one include a non-trivial central term in a Carrollian structure in higher dimensions?

To address this, we first revisit the origin of the central charge in the Bargmann algebra, following the derivation provided in Appendix \ref{gt}. The Galilei transformations \eqref{gemt} are obtained from the Lorentz transformations in the strict limit $c\to\infty$, corresponding to the Galilei algebra \eqref{Galileia}. Subsequently, by incorporating the leading $c$-dependent corrections to the Galilei transformations, presented in \eqref{gepmt}, and employing Newtonian mechanics, the mass parameter \eqref{gc} appears which in turn leads to the Bargmann algebra \eqref{Barga}. Therefore, two ingredients are required in this derivation of the Bargmann algebra: $(i)$ the inclusion of the leading $c$-dependent corrections, and $(ii)$ the use of Newtonian mechanics.

Analogously, to investigate a non-trivial central charge in the Carrollian case in higher dimensions, it is natural to involve two ingredients: $(i)$ the inclusion of the leading $c$-dependent corrections to the Carroll transformations, and $(ii)$ the use of post-Carrollian mechanics \cite{Najafizadeh:2025ksm} — the Carrollian counterpart of Newtonian mechanics. Following this procedure, we find that a central charge $M$ emerges, leading to the so-called ``Carroll–Bargmann algebra''. However, when the central charge vanishes, the Carroll–Bargmann algebra reduces to the so-called ``post-Carroll algebra'', rather than to the Carroll algebra\,\footnote{The term ``post-Carroll'' in this work differs from the terminology used in \cite{Ecker:2025ncp}. There, the parent algebra is the Carroll algebra, and the term is applied when the Hamiltonian and boost generators do not commute — a structure that does not yield a central charge in higher dimensions. In this work, however, we employ the term for our parent algebra, which is the ``post-Carroll algebra''. The case where the Hamiltonian and boost generators do not commute then yields a central charge, leading to a structure we call the ``Carroll–Bargmann algebra''.}\ensuremath{^{,}}\footnote{The leading $c$-dependent corrections to Carroll hydrodynamics are referred to as the {\it Carrollian regime} in \cite{Kolekar:2024cfg}.}.

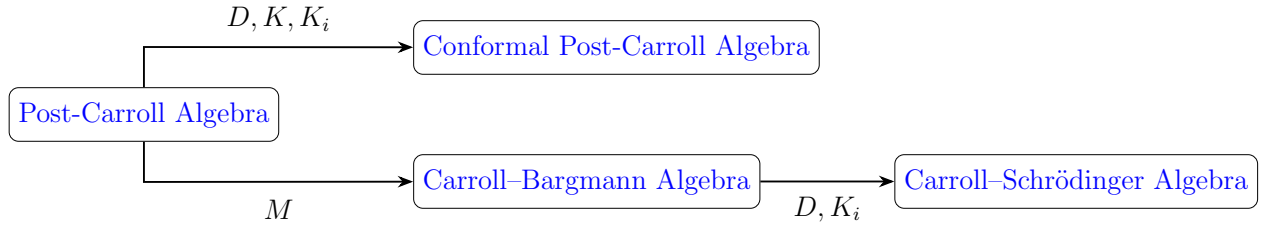
\begin{figure}[ht]
	\begin{center}
\resizebox{.95\textwidth}{!}{%
\begin{tikzpicture}[
	node distance=2cm,
	every node/.style={draw, rounded corners, align=center, minimum width=2.2cm, minimum height=0.8cm},
	labelnode/.style={draw=none, minimum width=2.2cm, minimum height=0.8cm},
	arrow/.style={-{Stealth}, thick}
	]
	
	\node (post) {\hyperref[pcarr]{Post-Carroll Algebra}}; 
	
	\node (conf-post) [right=of post, yshift=1cm] {\hyperref[confpcarr]{Conformal Post-Carroll Algebra}};
	
	\node (barg) [right=of post, yshift=-1cm] {\hyperref[pcarrb]{Carroll–Bargmann Algebra}};
	
	\node (schro) [right=of barg] {\hyperref[cpcarr]{Carroll–Schr\"odinger Algebra}};
	
	\draw[arrow] (post.north) -- ++(0, 0.6) -- node[above, labelnode] {$D,K,K_i$} (conf-post);
	
	\draw[arrow] (post.south) -- ++(0, -0.6) -- node[below, labelnode] {$M$} (barg);
	
	\draw[arrow] (barg) -- node[below, labelnode] {$D,K_i$} (schro);
	
\end{tikzpicture}
}	\end{center}	
	\caption{\small Extensions of the post-Carroll algebra with the generators $M$, $D$, $K$, $K_i$.}
	\label{fig}
\end{figure}

Therefore, we observe that while the Carroll algebra itself does not admit a non-trivial central charge in higher dimensions, an alternative Carrollian structure — namely, the post-Carroll algebra — does. Starting from the post-Carroll algebra, we construct its conformal extension by including the dilatation generator $D$ together with the special conformal transformation generators, temporal $K$ and spatial $K_i$. This yields the so-called ``conformal post-Carroll algebra'' with critical exponent $z=1$. In a similar fashion, we conformally extend the Carroll–Bargmann algebra  by including $D$ and $K_i$, leading to the so-called ``Carroll–Schr\"odinger algebra'' with $z=1/2$. This algebra is found to be the symmetry algebra of the Carroll–Schr\"odinger theory \cite{Najafizadeh:2024imn} in any dimension, which is a main result of this work. The parent algebra (i.e., the post-Carroll algebra), its central extension (the Carroll–Bargmann algebra), and the conformal extensions of both are depicted in Fig. \ref{fig}. 

We subsequently construct field theories invariant under each of these algebras. We find that the corresponding theories must involve complex fields. In other words, the post-Carroll algebra, and its extensions, do not constitute a symmetry for field theories of real fields. As we will see, this is due to the existence of the ``radial direction generator'' in the post-Carroll algebra. Such a generator is absent in the Carroll algebra, which allows for field theories involving both real and complex fields. 

Finally, considering the conformal post-Carroll algebra, we constrain the two-point functions between two complex scalar fields in a post-Carrollian conformal field theory (CFT). As we will see, the two-point functions exhibit both electric and magnetic sectors in 1+1 dimensions, whereas in higher dimensions only the magnetic sector contributes.

The layout of this paper is as follows. In Section \ref{pctra}, we briefly review Carroll transformations and the corresponding algebra. Subsequently, we present the necessary relations in post-Carrollian mechanics. Afterwards, we incorporate the leading $c$-dependent corrections to the Carroll transformations and, by applying post-Carrollian mechanics, derive the post-Carroll transformations. In Section \ref{beyond}, we use these transformations to derive the post-Carroll algebra and its central extension. Section \ref{conexten} presents conformal extensions, while consistent field theories are developed in Section \ref{ft}. Finally, in Section \ref{tpf}, we derive the general form of two-point functions in a post-Carrollian CFT, and we conclude in Section \ref{concl}. The appendices contain additional material making the paper self-contained: Appendix \ref{useful} provides useful commutation relations; using them, one can conveniently check algebras and invariance of actions. The approach we apply for the derivation of the Bargmann algebra is outlined in Appendix \ref{gt}, and Appendix \ref{Apct} details the derivation of the post-Carroll transformations using post-Carrollian mechanics.

\paragraph{Conventions:}
We use the mostly plus signature for the Minkowski metric $\eta_{\mu\nu}\equiv \text{diag} (-1,+1,\ldots,+1)$. We work in $1+d$ spacetime dimensions, where $d$ denotes the number of spatial dimensions. Small Latin indices $i$, $j$, $k$, $\dots$ therefore run over spatial coordinates: $1,\ldots,d$. We sometimes denote the contraction of spatial indices by $\vec a \cdot \vec b := a^i \,b_i$. We utilize $\vec v$ to denote the velocity of the object and $\vec u$ to denote the relative velocity between two frames. We use $\nabla_i$ as an alternative representation for the space translation generator \eqref{mrpc} and denote the standard gradient $\bm{\nabla}_i$ in bold to distinguish. Throughout the paper, we adopt the shorthand notations $\p_t:=\p/\p t$, $\p_x:=\p/\p x$, and $\p_i:=\p/\p x^i$. We define the Hermitian conjugation rules as
\be (\p_t)^\dagger\equiv-\,\p_t\,,\qquad (\p_{i})^\dagger\equiv-\,\p_{i}\,, \qquad(t)^\dagger\equiv t\,, \qquad (x_i)^\dagger\equiv x_i\,. \label{hcrules}
\ee

\section{Carroll transformations and beyond} \label{pctra}
To go beyond the Carroll transformations, let us first briefly review them by deriving their form from the Lorentz transformations. Consider a reference frame $S\,'$ moving with velocity $\vec{u}$ relative to the system $S$. The Lorentz transformations for energy and momentum are given by
\be 
E^{\,'}=\g\,\le(E-\vec{u}\cdot\vec{p}~\ri)\,,\qquad\quad
\vec{p}_{\,\shortparallel}^{~'}=\g\le(\vec{p}_{\,\shortparallel}-\vec{u}~\frac{E}{c^2}\ri)  \,,\qquad \quad
\vec{p}_{\!_\perp}^{~'}=\vec{p}_{\!_\perp}\,,\qquad\quad
\g=\frac{1}{\sqrt{1-\frac{u^2}{c^2}}}\,, \label{ltem}
\ee
where $\g$ is the Lorentz factor, $u^2=\vec{u}\cdot\vec{u}$, and momentum vector $\vec{p}=\vec{p}_{\,\shortparallel} \,+\, \vec{p}_{\!_\perp}$ is decomposed into components parallel and perpendicular to $\vec{u}$, so that $\vec{p}_{\,\shortparallel}\cdot\vec{p}_{\!_\perp}=0$, $\vec{u}\cdot\vec{p}=\vec{u}\cdot\vec{p}_{\,\shortparallel}$, and $\vec{u}\cdot\vec{p}_{\!_\perp}=0$. To get Carroll transformations, one first considers the non-relativistic case $u\ll c$, for which $\g\approx 1\,+\,u^2/(2\,c^2)$; this is why the seminal work \cite{Levy1965} identifies the Carroll transformations as the non-relativistic limit. However, the Carroll limit ($c\to 0$) is singular. To make the limit well-defined, one may introduce the Carroll boost parameter as $\vec{b}={\vec{u}}/{c^2}$. Substituting this and taking the strict limit $c\to 0$ (for which $\g\to 1$), the relations \eqref{ltem} transform into 
\begin{subequations}\label{emt}
	\begin{align}
		E^{\,'}&=E\,, \label{et}\\[3pt]
		\vec{p}_{\,\shortparallel}^{~'}&=\vec{p}_{\,\shortparallel}\,-\,\vec{b}\,E\,, \label{mt}\\[3pt] 
		{\vec{p}_{\!_\perp}}^{~'}&=\vec{p}_{\!_\perp}\,. \label{pt}
	\end{align}	
\end{subequations}
These are known as the Carroll transformations for energy and momentum \cite{deBoer:2021jej}\,\footnote{The transformations in \eqref{emt} are the counterparts of the Galilei transformations in \eqref{gemt}, obtained from the Lorentz transformations \eqref{ltem} in the strict limit $c\to\infty$.}. The relation \eqref{et} implies that energy remains invariant under a Carroll boost, as reflected by the vanishing commutator $[\,H\,,\,B_i\,] = 0$, where $H$ is the Hamiltonian and $B_i$ the Carroll boost generator. In addition, the relation \eqref{mt} reveals that momentum transforms under a boost when it is parallel to the boost direction, while the relation \eqref{pt} shows that the perpendicular part is boost-invariant. This behavior is encoded in the commutator $[\,P_i\,,\,B_j\,] = \delta_{ij}\,H$, where $P_i$ is the space translation generator. When rotational symmetry is included, the space translation and boost transform as vectors under spatial rotations \( J_{ij} \). Therefore, the set of generators \(\{ H, P_i, B_i, J_{ij} \}\), represented in differential form 
\be 
H=\p_t\,,\qquad\qquad P_i=\p_i \,, \qquad\qquad  B_i=x_i\,\p_t\,,\qquad\qquad  J_{ij} = x_i\,\p_j-x_j\,\p_i \,, \label{cargen}
\ee 
defines the Carroll algebra \cite{Levy1965,SenGupta:1966qer}, denoted by $\mathfrak{carr}(d+1)$, with the non-zero commutation relations
\begin{align}
	[\,P_i\,,\,B_j\,]=\d_{ij}\,H\,,\qquad
	[\,P_i\,,\,J_{jk}\,]=\d_{i[j}\,P_{k]}\,,\qquad[\,B_i\,,\,J_{jk}\,]=\d_{i[j}\,B_{k]}\,,\qquad[\,J_{ij}\,,\,J_{kl}\,]=\d_{[i[k}J_{l]j]}\,. \label{carra}
\end{align} 
Here, the Hamiltonian \( H \) is a central charge, commuting with all generators. However, as we will show later, we can introduce a non-vanishing commutator \([\,H\,,\,B_i\,] \ne 0\), breaking the Hamiltonian's role as a central charge. The price to pay is the inclusion of $c$-dependent corrections in the Carroll transformations \eqref{emt}. However, the presence of such corrections indicates that a post-Carrollian mechanical description must be employed \cite{Najafizadeh:2025ksm}. Thus, before including corrections to the Carroll transformations, let us first briefly review post-Carrollian mechanics in Subsec. \ref{pcme} and then investigate post-Carroll transformations in Subsec. \ref{pct}. Appendix \ref{gt} provides a similar approach for the Galilean case, which is useful for comparison with our results in this section.

\subsection{Post-Carrollian mechanics} \label{pcme}

Magnetic Carroll particles are always in motion, characterized by vanishing energy and non-vanishing momentum given by $E_{\c}=0$ and $\vec{p}_{\c}=mc\,{\hat{v}}$ \cite{deBoer:2021jej}\,\footnote{The other type, electric Carroll particles, are by contrast localized and fixed in space \cite{deBoer:2021jej}.}. Including the leading $c$-dependent corrections to these quantities gives rise to ``post-Carrollian mechanics''\,\footnote{The framework could be called ``post-magnetic Carrollian mechanics'', but for simplicity the term {\it magnetic} was omitted in \cite{Najafizadeh:2025ksm}, and the framework was referred to simply as ``post-Carrollian mechanics''. Accordingly, the particles under consideration were called ``post-Carroll particles'' instead of ``post-magnetic Carroll particles''.}. With these corrections included, one therefore deals with post-Carroll particles rather than magnetic Carroll particles. This framework has been explored in detail in \cite{Najafizadeh:2025ksm}, providing a mechanical description analogous to Newtonian mechanics.

As derived in \cite{Najafizadeh:2025ksm}, the total momentum in post-Carrollian mechanics is given by
\be 
\vec{p}~=~\vec{p}_{\c}\,+\,\vec{p}_{\pc}\,, \label{tp}
\ee 
which is the sum of two terms: the momentum $\vec{p}_{\c}$ of a magnetic Carroll particle \cite{deBoer:2021jej}
\begin{align}
		\vec{p}_{\c}&=mc~{\hat{v}}\,,  \label{cm} 
\end{align} 
and the momentum $\vec{p}_{\pc}$ of a post-Carroll particle. The energy $E_{\pc}$ and momentum $\vec{p}_{\pc}$ for a post-Carroll particle of mass $m$ are given by \cite{Najafizadeh:2025ksm}
\begin{align}
	E_{\pc}&=\frac{m\,c^3}{v}\,,\label{pcen}\\[8pt]
	\vec{p}_{\pc}&=\frac{m\,c^{\,3}}{2v^{\,2}}~{\hat{v}}\,, \label{pcm}
\end{align}
where $v=|\vec{v}|$ is the magnitude of the velocity vector, ${\hat{v}}=\vec{v}/v$ is the unit velocity vector, and $c$ denotes the speed of light. Combining the energy \eqref{pcen} and the magnitude of momentum $|\vec{p}_{\pc}|$ \eqref{pcm} yields the post-Carrollian energy-momentum relation
\be 
|\vec{p}_{\pc}|=\frac{E_{\pc}^{\,2}}{2m\,c^{\,3}}\,. \label{pcem}
\ee 
As observed, in the strict limit $c\to 0$, the leading $c$-dependent corrections vanish, so $E_{\pc}=0$ and $\vec{p}_{\pc}=0$. Thus, keeping $mc$ finite in the limit, we are left with the energy and momentum of magnetic Carroll particles, as expected.

We note that the relations \eqref{tp}–\eqref{pcem} in the post-Carrollian framework are the counterparts of \eqref{total mom}–\eqref{emrN} in the Newtonian mechanics. In particular, while Newtonian mechanics deals with total energy \eqref{total mom}, the post-Carrollian framework deals with total momentum \eqref{tp}. Thus, the ``rest energy'' in Newtonian mechanics \eqref{e0} plays a similar role to the momentum of a magnetic Carroll particle \eqref{cm} in post-Carrollian mechanics, a quantity that was accordingly called the ``rest momentum'' in \cite{Najafizadeh:2025ksm}. More details of this framework can be found in \cite{Najafizadeh:2025ksm}. Equipped with the quantities reviewed above, we now proceed to find post-Carroll transformations.

\subsection{Post-Carroll transformations} \label{pct}

To find post-Carroll transformations, we first include the leading $c$-dependent corrections to the Carroll transformations \eqref{emt}. To this end, we again consider the Lorentz transformations \eqref{ltem} and expand $\gamma$ in powers of $c$. Substituting the Carroll boost parameter $\vec{b}={\vec{u}}/{c^2}$, we obtain $\g=1\,+\,\tfrac{1}{2}\,b^2\,c^2\,+\,\mathcal{O}(c^4)$, where $b^2=\vec b\cdot\vec b$. Accordingly, the leading $c$-dependent corrections to the Carroll transformations, which may be referred to as ``expanded Carroll transformations'', are found to be
\bes{epmt}
\begin{align}	
	E^{\,'} &\,=\, E \,-\, c^2\,\vec{b}\cdot\vec{p}\,, \label{1epmt}\\[8pt]
	\vec{p}_{\,\shortparallel}^{~'} &\,=\, \vec{p}_{\,\shortparallel} \,-\, \vec{b}\,E \,+\, \tfrac{1}{2}\,c^2\,b^2\,\vec{p}_{\,\shortparallel}\,,  \label{2epmt}~~~~~~\\[8pt] 
	{\vec{p}_{\!_\perp}}^{~'} &\,=\, \vec{p}_{\!_\perp}\,. \label{3epmt}
\end{align} 
\ees
As expected, these reduce to the standard Carroll transformations \eqref{emt} in the strict limit $c \to 0$. We note that by introducing $\vec{p}_{\,\shortparallel}=\big({\,\vec{p}\,\cdot\,\vec{b}\,}/{b^2}\big)\,\vec{b}$, the momentum transformations, \eqref{2epmt} and \eqref{3epmt}, take the following compact form, which will be used in Appendix \ref{Apct}:
\be
\vec{p}^{~'}\,=~\vec{p}~-~\vec{b}\,E~+~\tfrac{1}{2}\,c^2\big(\vec{p}\cdot\vec{b}\,\big)\,\vec{b}\,.\label{compact}
\ee
We now employ post-Carrollian mechanics. This involves substituting the post-Carrollian energy \eqref{pcen} and the total momentum \eqref{tp} — with their relations given by \eqref{cm} \& \eqref{pcm} — into the expanded Carroll transformations \eqref{epmt}. Keeping only the leading terms, we obtain the ``post-Carroll transformations'' for energy and momentum 
\bes{epmtp}
\begin{empheq}[box={\fcolorbox{black}{gray!20}}]{align}
	~~~~~~~\phantom{\bigg(}E_{\pc}^{\,'} &\,=\, E_{\pc} \le(1-\vec{b}\cdot\vec{v}\,\ri)\,, \label{1epmtp}\\[8pt]
	\vec{p}_{_\mathrm{pc}}^{~'} &\,=\, \vec{p}_{_\mathrm{pc}}\le(1-\vec{b}\cdot\vec{v}\,\ri)^2\,. \label{2epmtp}\phantom{\bigg)}~~~~~~
\end{empheq} 
\ees
The details of derivation are provided in Appendix \ref{Apct}. We note that substituting the explicit form of the energy \eqref{pcen} and momentum \eqref{pcm} into the right-hand side of \eqref{epmtp} makes the $c$-dependence of these transformations manifest. In addition, these transformations \eqref{epmtp} demonstrate that under a Carroll boost, the post-Carrollian energy $E_{\pc}$ and momentum $\vec{p}_{_\mathrm{pc}}$ transform as rescalings. This rescaling behavior is further supported by the transformation of the velocity under a Carroll boost
\be 
\vec{v}^{~'}=\frac{\vec{v}}{~1-\vec{b}\cdot\vec{v}~\,}\,,\label{vt}
\ee 
as derived in \cite{deBoer:2021jej}. We note that this velocity transformation could also be obtained directly from \eqref{1epmtp} upon using \eqref{pcen}. Therefore, by applying the post-Carroll transformations \eqref{epmtp} together with the velocity transformation \eqref{vt}, we can conveniently verify that the post-Carrollian formulas given in \eqref{pcen}, \eqref{pcm}, and \eqref{pcem} transform covariantly. For instance, the post-Carrollian energy \eqref{pcen} indeed transforms covariantly; that is
\be 
E_{\pc}^{\,\,'}=\frac{m\,c^3}{v{\,'}} \qquad \longleftrightarrow \qquad E_{\pc}=\frac{m\,c^3}{v}\,. \label{cova}
\ee 
For comparison with the Newtonian framework, it is worth noting that the transformations in \eqref{epmtp} and \eqref{vt} under the post-Carrollian framework are the counterparts of those in \eqref{gs1} and \eqref{v}.

\section{Beyond the Carroll algebra} \label{beyond}

As the Carroll algebra \eqref{carra} follows from the Carroll transformations \eqref{emt}, we aim to go beyond and analogously construct an algebra corresponding to the post-Carroll transformations \eqref{epmtp}. In the Carroll case, the invariance of energy under boost, $E^{\,'}=E$ \eqref{et}, implies the vanishing commutator $[\,H\,, \,B_i\,] = 0$. In the post-Carroll regime, however, it finds that energy transforms as \eqref{1epmtp}, which can lead to a non-vanishing commutator $[\,H\,, \,B_i\,] \neq 0$.

To demonstrate this, we substitute the post-Carrollian energy \eqref{pcen} into the energy transformation \eqref{1epmtp}, which yields $E_{\pc}^{\,'} = E_{\pc} - c^3\,\vec{b}\cdot\hat{v}\,m$. Since $\vec b\cdot\hat{v}_{\!_\perp}=0$, we have $\vec b\cdot\hat v=\vec b\cdot\hat{v}_{\,\shortparallel}$\,. By a suitable spatial rotation, one can always choose a reference frame in which the boost vector aligns with the particle's position vector $\vec x$. In this frame, the unit radial vector $\vec{n}={\vec{x}}/{|\vec{x}|}$ is therefore parallel to the boost vector and, consequently, to $\hat{v}_{\,\shortparallel}$ provided the motion is purely radial. Hence, up to an overall sign, $\hat{v}_{\,\shortparallel}=\vec n$. Taking this into account, the energy transformation \eqref{1epmtp} simplifies to (setting $c=1$)
\be 
E_{\pc}^{\,'} = E_{\pc} - \,{b}^i\,{n}_i\,m\,, \label{ee'}
\ee
where $b^i$ and ${n}_i$ are the components of the boost vector $\vec{b}$ and the unit radial vector $\vec{n}$. This transformation can be encoded in the commutator between the Hamiltonian $H$ and the boost $B_i$ generators as\,\footnote{The counterpart of \eqref{ee'} in the Bargmann case is the pair \eqref{3gepmt} and \eqref{gc}. Together, these yield \eqref{M}, which is the counterpart of \eqref{MN}.}
\be
[\,H\,,\,B_i\,]=M\,n_i\,.\label{MN}
\ee
Here, we refer to $M$ as a central charge, even though it does not appear alone on the right-hand side. We use this terminology in the sense that $M$ actually commutes with all generators, as will be shown later. In addition, a new generator $n_i$ — which one may refer to as the ``radial direction generator'' — accompanies the central charge and is represented by
\be 
n_i=\frac{x_i}{r}\,, \label{N}
\ee 
where $r=|\vec{x}|=\sqrt{x^ix_i}$ denotes the magnitude of the position vector, so that $n^in_i=1$. Similarly, the momentum transformation \eqref{2epmtp}, expanded to first order in the boost parameter, yields \eqref{lo}. Using \eqref{pcen} and \eqref{pcm}, this leads to \eqref{1order}: $\vec{p}_{\pc}^{~'}=\vec{p}_{\pc}\,-\,E_{\pc}\,\hat{v}\,(\hat{v}\cdot\vec{b}\,)$. In component form, and under the previous assumption of radial motion, this becomes
\be 
p_{\pc}^{\,i}{}^{\!\!'} ~=~ p_{\pc}^{\,i} ~-~\,E_{\pc}\,{n}^i\,n^j\,{b}_j\,, \label{pcnne}
\ee 
which in turn demonstrates the commutator between the space translation $P_i$ and the boost $B_i$ as
\be 
[\,P_i\,,\,B_j\,]=n_i\,n_j\,H\,. \label{pcnn}
\ee 
Therefore, we find that the post-Carroll transformations \eqref{epmtp} correspond to an algebraic structure given by the commutation relations in \eqref{MN} and \eqref{pcnn}. If the central charge $M$ is set to zero, we are left with \eqref{pcnn} alone. Hence, by incorporating the spatial rotations $J_{ij}$, we introduce two algebras: 
\begin{itemize}
	\item {\bf Post-Carroll algebra}, with generators $\{H, P_i, B_i, J_{ij}, n_i\}$\,, (Subsec. \ref{eca});
	
	\item {\bf Carroll–Bargmann algebra}, with generators $\{H, P_i, B_i, J_{ij}, n_i, M\}$\,, (Subsec. \ref{pca}).
\end{itemize}

\subsection{Post-Carroll algebra: $\mathfrak{pcarr}(d+1)$} \label{eca}

We introduce the post-Carroll algebra by employing the same generators as the Carroll algebra \eqref{cargen}, with the difference that the generator set must be extended by $n_i$ and the space translation generator $P_i$ must take an alternative representation in order to satisfy \eqref{pcnn}. To this end, we find that the space translation generator should be represented as 
\begin{align} 
	P_i~&=~\nabla_i\,,  \label{mrpc}
\end{align}
where
\be 
\nabla_i\,:=\,\frac{n_i}{\,r\,}\le(\, \vec x\cdot\vec\partial_x~+~\frac{d-1}{2}\,\ri)\,, \label{nabla}
\ee
with $n_i$ defined in \eqref{N} and $d$ denoting the spatial dimension\,\footnote{\label{sg}We note that the operator \eqref{nabla}, denoted by $\nabla_i$, differs from the standard gradient operator. To distinguish between them, we denote the latter in bold as $\bm{\nabla}_i=\partial_i$ when necessary. For comparison, consider these two operators in spherical coordinates in $d=3$. The former \eqref{nabla} becomes $\vec\nabla=\hat r\le(\partial_r+\frac{1}{r}\ri)$, meaning that it is purely radial, while the standard gradient is $\vec{\bm{\nabla}}=\hat r\, \partial_r+\frac{\hat \theta}{r}\,\partial_\theta+\frac{\hat \phi}{r\sin{\theta}}\,\partial_\phi$. Thus, the two operators share the same radial part up to the additional term $\hat r/r$.}. This operator was first introduced in Appendix B of \cite{Najafizadeh:2025ksm}; for further details, see Section \ref{cst}, especially \eqref{prisc}. It is defined to be anti-Hermitian, $(\nabla_i)^{\dagger}=-\,\nabla_i$, which necessitates the inclusion of the second term in \eqref{nabla}. Accordingly, using \eqref{mrpc}, we represent the generators of the post-Carroll algebra \{\( H \), \( P_i \), \( B_i \), $n_i$, \( J_{ij} \)\} by
\beak{lll}\phantomsection\label{ecarrg}
H=\p_t\,,\qquad\quad & P_i={\nabla}_i\,, \qquad\quad &  B_i=x_i\,\p_t\,,\qquad\quad n_i= \frac{x_i}{r}\,,\qquad\quad J_{ij} = x_i\,\p_j-x_j\,\p_i\,. \nk 
\eeak
These generators are taken to be anti-Hermitian, $O^\dagger=-\,O$, by applying the Hermitian conjugation rules \eqref{hcrules}, with the exception of $n_i$, which is assumed Hermitian for simplicity, i.e., $n_i^\dagger=n_i$. While this choice is irrelevant at the level of algebra, it becomes significant in field theory, where $N_i=i\,n_i$ must be defined to be anti-Hermitian in order to be properly identified as a symmetry of the action, as will be shown later. We find that the set of generators in \eqref{ecarrg} satisfies the following non-zero commutation relations (see Appendix \ref{useful} for useful relations)
\be 
\fcolorbox{black}{gray!20}{$~~~~
	\begin{aligned}\phantom{\bigg(}
		&[\,P_i\,,\,B_j\,]=n_i\,n_j\,H\,,\qquad &&[\,P_i\,,\,J_{jk}\,]=\d_{i[j}\,P_{k]}\,,\qquad &&[\,J_{ij}\,,\,J_{kl}\,]=\d_{[i[k}J_{l]j]}\,, \\ 
		&[\,n_i\,,\,J_{jk}\,]=\d_{i[j}\,n_{k]}\,,\qquad &&[\,B_i\,,\,J_{jk}\,]=\d_{i[j}\,B_{k]}\,.\qquad && \label{pcarr} 
		\phantom{\bigg(}
	\end{aligned}
	~~~~$}
\ee
We refer to this structure as the ``post-Carroll algebra'' in arbitrary spacetime dimensions and denote it by $\mathfrak{pcarr}(d+1)$. In this algebra, $n_i$ transforms as a vector under rotations, just like $P_i$ and $B_i$, while the Hamiltonian $H$ is a central charge in the sense explained below \eqref{MN}. It is understood that any commutator of the operators \eqref{ecarrg} automatically satisfies the Jacobi identity. Nevertheless, when the algebra \eqref{pcarr} is considered in its abstract form — i.e., without any explicit representation of the generators given in \eqref{ecarrg} — we can conveniently verify that the Jacobi identity holds.

We note that in the absence of rotation generators $J_{ij}$, the post-Carroll algebra \eqref{pcarr} reduces to \eqref{pcnn}. This implies
\be 
\fcolorbox{black}{gray!20}{$~~~~
	\begin{aligned}\phantom{\bigg(}
	[\,P_i\,,\,x_j\,]=n_i\,n_j\,, \label{ph} 
		\phantom{\bigg(}
	\end{aligned}
	~~~~$}
\ee
which we refer to as the ``post-Heisenberg algebra'' and denote by $\mathfrak{ph}_d$. Therefore, the key difference between the standard Heisenberg algebra $\mathfrak{h}_d$, $[\,P_i\,,\,x_j\,]=\delta_{ij}$, and the post-Heisenberg algebra lies in their respective representations: the former admits the representation $P_i=\partial_i$, whereas the latter requires $P_i=\nabla_i$, so the generator $n_i$ necessarily appears in the algebra \eqref{ph}.

In 1+1 spacetime dimensions, the rotation generators $J_{ij}$ vanish and $n_i$ becomes trivial, effectively reducing to unity. This indicates that $n_i$, much like the rotation generator $J_{ij}$, plays a nontrivial role only in spatial dimensions $d>1$. Moreover, in 1+1 dimensions, where $d=1$, the operator $\nabla_i$ reduces to the standard partial derivative $\partial_x$. Consequently, the algebra \eqref{pcarr} simplifies to $[\,P\,,\,B\,]=H$, which is precisely the Carroll algebra \eqref{carra}. Hence, $\mathfrak{pcarr}(1+1)\cong\mathfrak{carr}(1+1)$, demonstrating that in 1+1 dimensions, the Carroll algebra and the post-Carroll algebra are identical. Similarly, in $d=1$, the Heisenberg algebra and the post-Heisenberg algebra become identical as well: $\mathfrak{h}_1\cong\mathfrak{ph}_1$.

\subsection{Carroll–Bargmann algebra: $\mathfrak{carrb}(d+1)$} \label{pca}

As previously mentioned, the Carroll algebra \eqref{carra} does not admit a non-trivial central charge $M$ in spatial dimensions $d>1$. In contrast, the post-Carroll algebra \eqref{pcarr} accommodates a central charge $M$ in arbitrary dimensions, breaking the Hamiltonian's role as a central charge. Accordingly, by including $M$, we extend the post-Carroll generators \eqref{ecarrg} to \{\( H \), \( P_i \), \( B_i \), \( J_{ij} \), $n_i$, $M$\}, represented by
\beak{lll}\phantomsection\label{pcarrg}
H=\p_t\,,\qquad\quad & P_i=\nabla_i \,, \qquad\quad &  B_i=x_i\,\p_t\,+\,t\,Mn_i\,,\qquad\quad  J_{ij} = x_i\,\p_j-x_j\,\p_i\,, \\[8pt]
n_i= \frac{x_i}{r}\,,\qquad\quad & M=-\,im\,. \nk 
\eeak
Here, we represented the central charge $M$ as anti-Hermitian to preserve the anti-Hermiticity of the boost generator $B_i$. We find that these generators satisfy the following non-zero commutation relations
\be 
\fcolorbox{black}{gray!20}{$~~~~
	\begin{aligned}\phantom{\bigg(}
		&[\,P_i\,,\,B_j\,]=n_i\,n_j\,H\,,\qquad &&[\,P_i\,,\,J_{jk}\,]=\d_{i[j}\,P_{k]}\,,\qquad &&[\,J_{ij}\,,\,J_{kl}\,]=\d_{[i[k}J_{l]j]}\,, \\ 
		&[\,n_i\,,\,J_{jk}\,]=\d_{i[j}\,n_{k]}\,,\qquad &&[\,B_i\,,\,J_{jk}\,]=\d_{i[j}\,B_{k]}\,,\qquad &&[\,H\,,\,B_i\,]=M\,n_i\,. \label{pcarrb} 
		\phantom{\bigg(}
	\end{aligned}
	~~~~$}
\ee
We refer to this as the ``Carroll–Bargmann algebra'' in arbitrary dimensions, and denote it by $\mathfrak{carrb}(d+1)$. As observed here, this algebra implies that $M$ is a central charge commuting with all generators, while the Hamiltonian $H$ is no longer central. In 1+1 spacetime dimensions, the algebra \eqref{pcarrb} reduces to 
\begin{align} 
[\,H\,,\,B\,]=M\,, \qquad\qquad  [\,P\,,\,B\,]=H\,. \label{cb}
\end{align}
This was called the Carroll–Bargmann algebra and denoted $\mathfrak{carrb}(1+1)$ in \cite{Najafizadeh:2024imn}, which motivated us to adopt the same name for \eqref{pcarrb} in higher dimensions. Furthermore, in 1+1 dimensions, the transformations \eqref{ee'} and \eqref{pcnne}, derived from the post-Carroll transformations \eqref{epmtp}, simplify respectively to
\begin{align}  
E_{\pc}{\!\!\!}' ~=~ E_{\pc} - ~{b}\,m\,,\qquad\qquad p_{\pc}{\!\!\!\!}'~=~ p_{\pc} -~ {b}\,E_{\pc}\,. \label{th}
\end{align} 
These transformations correspond precisely to the Carroll–Bargmann algebra \eqref{cb}, confirming once again that the inclusion of leading $c$-dependent corrections to the Carroll transformations \eqref{emt} results in $M$ becoming a central charge while breaking the Hamiltonian's role as one.

\section{Conformal extensions} \label{conexten}

In this section, we investigate the conformal extension of the post-Carroll algebra \eqref{pcarr}, presented in Subsec. \ref{cpca}, as well as the conformal extension of the Carroll–Bargmann algebra \eqref{pcarrb}, discussed in Subsec. \ref{pcsha}. We note that these are not the only possible extensions; indeed, one may follow the procedure outlined in \cite{Afshar:2024llh} to systematically classify and identify all minimal conformal extensions of the post-Carroll algebra.

\subsection{Conformal post-Carroll algebra: $\mathfrak{cpcarr}(d+1)$} \label{cpca}
We observe that the post-Carroll algebra \eqref{pcarr} admits a conformal extension through the inclusion of the generators \{$D$, $K$, $K_i$\}, where $D$ generates dilatations, while $K_i$ and $K$ generate spatial and temporal post-Carrollian special conformal transformations, respectively. Consequently, the set of generators in \eqref{ecarrg} extends to \{$H$, $P_i$, $B_i$, $J_{ij}$, $n_i$, $D$, $K$, $K_i$\}, represented by
\beak{llll}\phantomsection
H=\p_t\,,~\qquad\quad & P_i=\nabla_i \,, ~\qquad\quad &  B_i=x_i\,\p_t\,,~\qquad\quad&  J_{ij} = x_i\,\p_j-x_j\,\p_i\,, \\[8pt]
n_i= \frac{x_i}{r}\,,~\qquad\quad &  D=t\,\p_t\,+\,x^i\p_i\,+\,\omega\,, ~\qquad\quad & K=x^2\,\p_t\,, ~\qquad\quad & K_i=2\,x_i\,D\,-\,x^2\,\nabla_i\,, \nk  \label{gcpca}
\eeak
where $\omega$ is the dilatation weight. We find that these satisfy the following non-zero commutation relations
\be 
\fcolorbox{black}{gray!20}{$~~~~
	\begin{aligned}\phantom{\bigg(}
		&[\,P_i\,,\,B_j\,]=n_i \,n_j\,H\,,\qquad && [\,D\,,\,H\,]=-\,H\,, \qquad && [\,K_i\,,\,H\,]=-\,2\,B_i\,,\\\phantom{\bigg(}
		&[\,P_i\,,\,J_{jk}\,]=\d_{i[j}\,P_{k]}\,,\qquad && [\,D\,,\,P_i\,]=-\,P_i\,, \qquad && [\,K_i\,,\,P_j\,]=-\,2\,n_i\,n_j\,D\,,\\\phantom{\bigg(}
		&[\,B_i\,,\,J_{jk}\,]=\d_{i[j}\,B_{k]}\,,\qquad && [\,D\,,\,K\,]=K\,, \qquad && [\,K_i\,,\,B_j\,]=-\,n_i\,n_j\,K\,,\\\phantom{\bigg(}
		&[\,n_i\,,\,J_{jk}\,]=\d_{i[j}\,n_{k]}\,,\qquad && [\,D\,,\,K_i\,]=K_i\,, \qquad && [\,K_i\,,\,J_{jk}\,]=\d_{i[j}\,K_{k]}\,,\\\phantom{\bigg(}
		&[\,J_{ij}\,,\,J_{kl}\,]=\d_{[i[k}J_{l]j]}\,,\qquad && [\,K\,,\,P_i\,]=-\,2\,B_i\,. &&\nk
		\phantom{\bigg(}\label{confpcarr}
	\end{aligned}
	~~~~$}
\ee
We refer to this structure as the ``conformal post-Carroll algebra'' in any dimension with the critical exponent $z=1$ and denote it by $\mathfrak{cpcarr}(d+1)$. Thus, we have the following hierarchy of subalgebras
\be 
\mathfrak{ph}_d ~\subset~\mathfrak{pcarr}(d+1)~\subset ~\mathfrak{cpcarr}(d+1)\,.
\ee 
In 1+1 spacetime dimensions, the conformal post-Carroll algebra \eqref{confpcarr} reduces to the two-dimensional Carrollian conformal algebra (CCA), denoted $\mathfrak{cca}(1+1)$, such that $\mathfrak{cpcarr}(1+1)\cong\mathfrak{cca}(1+1)$, as expected.

We note that the vector generators in \eqref{gcpca} are each proportional to the generator $n_i$. Therefore, for an abstract algebra in a post-Carrollian regime, we assume that each vector generator $O_i$ is proportional to $n_i$, i.e. $O_i=\alpha\,n_i$ for some $\alpha$. To illustrate how this works in practice, take into account the generators \{$P_i$, $B_i$, $K_i$\} of the algebra \eqref{confpcarr}. Imposing the Jacobi identity on these generators yields
$
[\,P_j\,,\,[\,B_k\,,\,K_i\,]]+[\,B_k\,,\,[\,K_i\,,\,P_j\,]]+[\,K_i\,,\,[\,P_j\,,\,B_k\,]]=2\,n_k\,(n_i\,B_j-n_j\,B_i)$, which vanishes identically once the property $B_i=\alpha\,n_i$ is used.

\subsection{Carroll–Schr\"odinger algebra: $\mathfrak{carrsch}(d+1)$} \label{pcsha}

Here, we aim to construct a conformal extension of the Carroll–Bargmann algebra \eqref{pcarrb}. We find that such an extension is possible by including the conformal generators \(D\) (dilatation with $z=1/2$) and \(K_i\) (spatial special conformal transformations) to the Carroll–Bargmann generators \eqref{pcarrg}. Therefore, the full set of generators is given by \{$H$, $P_i$, $B_i$, $J_{ij}$, $n_i$, $M$, $D$, $K_i$\}, with the following representations
\beak{llll}\phantomsection
H=\p_t\,,~\qquad & P_i=\nabla_i \,, ~\qquad &  B_i=x_i\,\p_t\,+\,t\,Mn_i\,,~\qquad&  J_{ij} = x_i\,\p_j-x_j\,\p_i\,, \nk  \label{cscgener}\\[8pt]
n_i= \frac{x_i}{r}\,,~\qquad & M=-\,i\,m\,, ~\qquad & D=t\,\p_t\,+\,2\,x^i\p_i\,+\,\omega\,,~\qquad & K_i=x_i\,D-x^2\,\nabla_i+\tfrac{1}{2}\,M\,t^2\,n_i\,,
\eeak
where $\omega$ denotes the dilatation weight. We find that these generators satisfy the following non-zero commutation relations
\be 
\fcolorbox{black}{gray!20}{$~~~~
	\begin{aligned}\phantom{\bigg(}
		&[\,P_i\,,\,B_j\,]=n_i\, n_j\,H\,,\qquad &&  [\,H\,,\,B_i\,]=M\,n_i\,,\qquad &&  [\,K_i\,,\,H\,]=-\,B_i\,,\\\phantom{\bigg(}
		&[\,P_i\,,\,J_{jk}\,]=\d_{i[j}\,P_{k]}\,,\qquad && [\,D\,,\,H\,]=-\,H\,,\qquad && [\,K_i\,,\,P_j\,]=-\,n_i\, n_j\,D\,,\\\phantom{\bigg(}
		&[\,B_i\,,\,J_{jk}\,]=\d_{i[j}\,B_{k]}\,,\qquad && [\,D\,,\,P_i\,]=-\,2P_i\,,\qquad && [\,K_i\,,\,J_{jk}\,]=\d_{i[j}\,K_{k]}\,,\\\phantom{\bigg(}
		&[\,n_i\,,\,J_{jk}\,]=\d_{i[j}\,n_{k]}\,,\qquad && [\,D\,,\,B_i\,]=B_i\,,\\\phantom{\bigg(}
		&[\,J_{ij}\,,\,J_{kl}\,]=\d_{[i[k}J_{l]j]}\,,\qquad && [\,D\,,\,K_i\,]=2K_i\,.\label{cpcarr} \phantom{\bigg(}
	\end{aligned}
	~~~~$}
\ee
We refer to this structure as the ``Carroll–Schr\"odinger algebra'' in arbitrary spacetime dimensions with critical exponent $z=1/2$ and denote it by $\mathfrak{carrsch}(d+1)$. Therefore, we have the following hierarchy of subalgebras
\be 
\mathfrak{ph}_d ~\subset ~ \mathfrak{pcarr}(d+1)~\subset ~\mathfrak{carrb}(d+1)~\subset ~\mathfrak{carrsch}(d+1)\,.
\ee
In 1+1 spacetime dimensions, where the rotation generator $J_{ij}$ vanishes and $n_i$ becomes trivial, the algebra \eqref{cpcarr} reduces to
\be\begin{aligned}
	&[\,P\,,\,B\,]=H\,, \quad &&[\,H\,,\,B\,]={M}\,, \quad &&[\,H\,,\,D\,]=H\,, \quad &&[\,P\,,\,D\,]=2\,P\,, \\[2pt]
	&[\,D\,,\,B\,]=B\,, \quad &&[\,D\,,\,K\,]=2\,K\,, \quad && [\,H\,,\,K\,]=B\,,\quad &&[\,P\,,\,K\,]=D\,.\label{csch}
\end{aligned} 
\ee 
This algebra was called the Carroll–Schr\"odinger algebra in \cite{Najafizadeh:2024imn,Afshar:2024llh}, denoted $\mathfrak{carrsch}(1+1)$. This motivated our naming for the algebra \eqref{cpcarr}, which is a higher-dimensional extension of \eqref{csch}. We note that, in the Carroll framework, it was shown that the algebra \eqref{csch} cannot be extended to higher dimensions \cite{Afshar:2024llh}. Here, however, we find that such an extension does exist, but it lies beyond the Carroll regime —  namely, in the post-Carroll setting. In addition, the terminology for \eqref{cpcarr} highlights a key similarity with the Schr\"odinger algebra, which includes the conformal generators \{$D$, $K$\} with $z=2$ (see, e.g., \cite{Duval:2024eod,Boisvert:2025hex}). By analogy, the Carroll–Schr\"odinger algebra \eqref{cpcarr} contains the conformal generators \{$D$, $K_i$\} with $z=1/2$.

\section{Field theories}  \label{ft}

In this section, we introduce post-Carrollian field theories that are invariant under the post-Carroll algebra. We then examine their invariance under the conformal post-Carroll algebra. In addition, we study the symmetries of the Carroll–Schr\"odinger field theory \cite{Najafizadeh:2024imn} in higher dimensions and show that it is invariant under the Carroll–Schr\"odinger algebra \eqref{cpcarr}.

\subsection{Post-Carrollian theory} \label{PCT}
In analogy with Carrollian theories (see e.g. \cite{deBoer:2021jej,Bergshoeff:2022qkx}), we identify two types of post-Carrollian theories: {\it electric}, where time derivatives dominate, and {\it magnetic}, where spatial derivatives dominate.
\paragraph{Electric:} In $1+d$ spacetime dimensions, we introduce the ``electric post-Carroll action'' as
\be 
\fcolorbox{black}{gray!20}{$~~
	\begin{aligned}\phantom{\Bigg(}
		S=-\,\int dt \,d^dx ~\phi^* \,\p_t^{\,2}\,\phi\,, \label{epca} \phantom{\Bigg(}
	\end{aligned}
	~~$}
\ee
where $\phi$ is a complex scalar field. An action with a real scalar field would break the post-Carroll algebra as a symmetry, as we will see. The invariance of the action \eqref{epca} under the generators $H$, $P_i$, $B_i$, and $J_{ij}$ given in \eqref{ecarrg} is evident, since these are anti-Hermitian operators and all commute with $\partial_t^{\,2}$. However, the action \eqref{epca}, even when formulated with a real scalar field, is not invariant under the generator $n_i$, as it is represented as Hermitian in \eqref{ecarrg}. To resolve this, we introduce the anti-Hermitian radial direction generator, via the imaginary unit $i$,
\be 
{N}_i=i\,n_i\,,  \label{Ni}
\ee
which satisfies $N_i^\dagger=-\,N_i$. In this way, the action \eqref{epca} becomes invariant under the transformation 
	\begin{align}
	\phantom{\Bigg(}
	\d\phi= \bigg(~\lambda_{_H}\,H~+~\lambda^i_{_P}\,P_i~+~\lambda^i_{_B}\,B_i~+~\lambda^i_{_N}\,(i\,n_i)~+~\lambda^{ij}_{_J}\,J_{ij}~\bigg)\,\phi\,,\label{transf}
	\phantom{\Bigg(}
	\end{align}
where $H$, $P_i$, $B_i$, $i\,n_i$, $J_{ij}$ are the generators of the post-Carroll algebra \eqref{pcarr}, represented in \eqref{ecarrg}, and $\lambda_{_H}$, $\lambda^i_{_P}$, $\lambda^i_{_B}$, $\lambda^i_{_N}$, $\lambda^{ij}_{_J}$ denote the corresponding transformation parameters. This invariance implies that the post-Carroll algebra \eqref{pcarr}, when expressed in terms of $N_i$, is precisely the symmetry algebra of the complex action \eqref{epca}.

From \eqref{transf}, we note that the field transformation under the generator $N_i$ alone is $\delta_{_N}\phi=\lambda^i_{_N}(i\,n_i)\,\phi$. This transformation, involving the imaginary unit $i$, implies that the field $\phi$ must be taken as a complex scalar, as initially considered in \eqref{epca}. Therefore, the presence of $n_i$ (and hence $N_i$) in the post-Carroll algebra \eqref{pcarr} allows it to be a symmetry for complex actions, such as \eqref{epca}; however, the algebra does not furnish a symmetry for real actions\,\footnote{By contrast, the Carroll algebra \eqref{carra} serves as a symmetry for both real and complex actions.}. Hence, from now on, we restrict our attention to complex actions. An exception arises in 1+1 spacetime dimensions, where the generator $n_i$ is trivially realised, allowing for the possibility of a real action as well.

\paragraph{Magnetic:} In $1+d$ spacetime dimensions, we present the ``magnetic post-Carroll action'' as
\be 
\fcolorbox{black}{gray!20}{$~~
	\begin{aligned}
		\phantom{\Bigg(}
	S=\int dt \,d^dx ~\Big(\,\chi^*\,\p_t\,\phi~-~\phi^*\,\p_t\,\chi~+~\phi^* \,\nabla_i\nabla^{\,i}\,\phi\,\Big)\,, \label{mpca}
		\phantom{\Bigg(}
	\end{aligned}
	~~$}
\ee
where $\phi$ is a complex scalar field, $\chi$ is a complex scalar Lagrange multiplier, and $\nabla_i$ is defined in \eqref{nabla}, with its square expanded in \eqref{nabnab}. The invariance under the anti-Hermitian generators $H$, $P_i$, $i\,n_i$, and $J_{ij}$ is manifest, as they all commute with both $\partial_t$ and $\nabla_i\nabla^{\,i}$ (see Appendix \ref{useful} for useful relations). This is demonstrated by commutation relations such as $[\,\nabla_i\nabla^{\,i}\,,\,n_j\,]=0$ and $[\,\nabla_i\nabla^{\,i}\,,\,J_{jk}\,]=0$. On the other hand, although the boost generator $B_i$ commutes with $\partial_t$, it does not commute with $\nabla_i\nabla^{\,i}$, as seen from the commutation relation $[\,\nabla_i\nabla^{\,i}\,,\,B_j\,]=2HP_j$. Therefore, similar to the magnetic Carroll case, the action \eqref{mpca} is not invariant under a standard boost transformation. Instead, it is invariant under the transformations
	\begin{align}
		\d_{_B}\,\phi&~=~\lambda^i_{_B}~B_i\,\,\phi\,,\\[8pt]
		\d_{_B}\,\chi&~=~\lambda^i_{_B}~B_i\,\,\chi~+~\lambda^i_{_B}\,\nabla_i\,\phi\,,\label{transfma2}
	\end{align}
where the second term in the latter transformation restores boost invariance. In this way, we find that the magnetic action \eqref{mpca} possesses the symmetries of the post-Carroll algebra \eqref{pcarr}. In 1+1 dimensions, where $\nabla_i$ reduces to $\p_x$, the action \eqref{mpca} (up to total derivatives) and the transformations \eqref{transfma2} simplify to the magnetic Carrollian theory \cite{deBoer:2021jej,Bergshoeff:2022qkx}, formulated for a complex field.

\subsection{Conformal post-Carroll theory}

Since the electric action \eqref{epca} and the magnetic one \eqref{mpca} possess the symmetries of the post-Carroll algebra \eqref{pcarr}, we now investigate whether they also encompass the symmetries of its conformal extension \eqref{confpcarr}. To this end, it is sufficient to examine the invariance of these actions under the conformal generators $D$, $K$, and $K_i$ \eqref{gcpca}. For this purpose, one also requires the Hermitian conjugates of these generators,
\be 
D^\dagger =-\,D+2\,\omega-d-1\,,\qquad\qquad  K^\dagger=-\,K\,,\qquad\qquad K_i^\dagger=-\,K_i+2\,(\,2\,\omega-d-1\,)\,x_i\,,
\ee  
which can be obtained by applying the Hermitian conjugation rules \eqref{hcrules} to the generators $D$, $K$, $K_i$. Accordingly, for the electric case, we find that the action \eqref{epca} is invariant under the transformation
\begin{align}
	\d\phi=\Big(~\lambda_{_D}\,D~+~\lambda_{_K}\,K~+~\lambda^i_{_K}\,K_i~\Big)\,\phi\,,\label{transfc}
\end{align} 
when the dilatation weight is set to $\omega=\frac{d\,-\,1}{2}$. Here, $\lambda_{_D}$, $\lambda_{_K}$, $\lambda^i_{_K}$ denote the transformation parameters associated with the generators $D$, $K$, $K_i$. Thus, the electric action \eqref{epca} exhibits the symmetries of the conformal post-Carroll algebra \eqref{confpcarr}. In addition, for the magnetic case, we find that the action \eqref{mpca} is invariant under the transformations  
\begin{align}
	\d\phi&=\Big(~\lambda_{_D}\,D~+~\lambda_{_K}\,K~+~\lambda^i_{_K}\,K_i~\Big)\,\phi\,,\\[4pt]
	\d\chi&=\Big(~\lambda_{_D}\,D~+~\lambda_{_K}\,K~+~\lambda^i_{_K}\,K_i~\Big)\,\chi~+~2\,\Big(x^i\,\lambda_{_K}~+~t\,\lambda^i_{_K}\Big)\,\nabla_i\,\phi\,,\label{transfcm}
\end{align} 
provided the dilatation weights are fixed to $\omega_\phi=\frac{d\,-\,1}{2}$ and $\omega_\chi=\frac{d\,+\,1}{2}$.

\subsection{Carroll–Schr\"odinger theory} \label{cst}

In $1+d$ spacetime dimensions, the Carroll–Schr\"odinger theory is given by the action (with $c=1=\hbar$) \cite{Najafizadeh:2024imn}
\be 
S=\int dt\,d^dx~\psi^\dagger\le(i\,\nabla_x~+~\frac{1}{2m}~\p_t^{\,2}\ri)\psi\,,\label{csa}
\ee
where $m$ is the mass parameter and the operator $\nabla_x$ is defined as
\be 
\nabla_x=\frac{1}{r}\le(\vec x\cdot\vec\partial_x~+~\frac{d-1}{2}\ri)\,. \label{nablax}
\ee 
Comparing this operator with that defined in \eqref{nabla}, we can obtain the relation between them, namely $\nabla_x=\vec n\cdot\vec\nabla$, or equivalently $\nabla_i=n_i\,\nabla_x$. Furthermore, as demonstrated in \cite{Najafizadeh:2025ksm}, one finds the identity
\be 
{\nabla_i\nabla^{\,i}}=(\nabla_x)^2\,, \label{identity}
\ee
whose explicit expanded form is given by \eqref{nabnab}. This identity can also be verified directly using the commutator $[\,\nabla_x\,,\,n_i\,]=0$. Varying the action \eqref{csa} with respect to the field $\psi^\dagger$ gives rise to the Carroll–Schr\"odinger equation\,\footnote{In $d>1$, it was initially referred to as the ``generalized Carroll–Schr\"odinger equation'' in \cite{Najafizadeh:2024imn} because its symmetry algebra was unknown. However, since this work identifies the symmetry algebra in arbitrary dimensions, called the ``Carroll–Schr\"odinger algebra'' \eqref{cpcarr}, we adopt the unified name ``Carroll–Schr\"odinger equation'' in all dimensions.}
\be 
\le(i\,\nabla_x~+~\frac{1}{2m}~\p_t^{\,2}\ri)\psi=0\,. \label{inany}
\ee
This equation was derived using two different approaches in references \cite{Najafizadeh:2024imn} and \cite{Najafizadeh:2025ksm}. In \cite{Najafizadeh:2024imn}, it was obtained from the Klein–Gordon equation, $\le(-\,\p_t^{\,2}\,+\,\p^{\,i}\p_{\,i}\,+\,m^2\,\ri)\phi=0$, of a complex tachyon field, where units with $c=1=\hbar$ are used. Applying the field redefinition $\phi=\frac{1}{\sqrt{m}}\,\exp(-\,im\,r)\,\psi$, results in $\le(-\,\p_t^{\,2}\,+\,\p^{\,i}\p_{\,i}\,-\,2\,im\nabla_x\,\ri)\psi=0$. The derivation then proceeds by rescaling the mass parameter $m\to m/\epsilon^2$, and taking the Carrollian limit: $x_i\to x_i$, $t\to\epsilon \,t$, $\epsilon\to 0$. 

Another derivation, presented in Appendix B of \cite{Najafizadeh:2025ksm}, begins with the post‑Carrollian energy-momentum relation \eqref{pcem}, which for $c=1$ becomes $|\vec{p}_{\pc}|=(\,{{p}^{\,i}_{\pc}\,{p}_{\,i}^{\,\pc}})^{1/2}={E_{\pc}^{\,2}}/{2m}$. Applying the following quantum mechanical prescription
\be
{p}^{\,i}_{\pc} ~\longrightarrow~ i \,\nabla^{\,i}~, \quad\qquad {E}_{\pc}~\longrightarrow~ -\,i\,\p_t\,,\label{prisc}
\ee 
with $\nabla_i$ defined in \eqref{nabla}, yields the Carroll–Schr\"odinger equation \eqref{inany}, where the identity \eqref{identity} is used. It is worth noting that the representation of space translation generator we considered in \eqref{mrpc} originates from the prescription \eqref{prisc}, introduced in \cite{Najafizadeh:2025ksm}.

In 1+1 dimensions, the operator $\nabla_x$ simplifies to the usual partial derivative $\p_x$, and thus the equation \eqref{inany} reduces to $\le(i\,\partial_x~+~\frac{1}{2m}~\p_t^{\,2}\ri)\psi=0$, which is isomorphic to the standard Schr\"odinger equation upon exchanging $x$ and $t$. This equation, and further correspondences with the Schr\"odinger equation, has been studied in quantum systems in several recent works \cite{Casanova:2025skc,Rojas:2025rot,Rojas:2025ygg,Rojas:2026rhf}.

Although the symmetries of the action \eqref{csa} in 1+1 spacetime dimensions were presented in \cite{Najafizadeh:2024imn}, they have not been identified in higher dimensions. Here, we show that the Carroll–Schr\"odinger algebra \eqref{cpcarr} obtained in this work is indeed the symmetry of the Carroll–Schr\"odinger action \eqref{csa} in arbitrary dimensions. To establish this, it is sufficient to demonstrate the invariance of the action \eqref{csa} under the generators given in \eqref{cscgener}, as shown below. We begin by introducing the ``Carroll–Schr\"odinger operator''
\be 
\mathbb{K}=i\,\nabla_x~+~\frac{1}{2m}~\p_t^{\,2}\,, \label{cso}
\ee 
and consider the field transformation as
\be 
\delta\psi= \lambda_{_\mathcal{O}}~\mathcal{O}\,\psi\,,
\ee 
where $\mathcal{O}$ denotes any of the generators \{$H$, $P_i$, $B_i$, $J_{ij}$, $i\,n_i$, $M$, $D$, $K_i$\} of the Carroll–Schr\"odinger algebra, represented in \eqref{cscgener}, and $\lambda_{_\mathcal{O}}$ is the corresponding transformation parameter. Since $\delta\psi^\dagger=\lambda_{_\mathcal{O}}\,\psi^\dagger\,\mathcal{O}^\dagger$, the invariance of the action \eqref{csa} under $\mathcal{O}$ becomes   
\be 
\delta_{_\mathcal{O}} S=\int dt\,d^dx\,\lambda_{_\mathcal{O}}\,\psi^\dagger\le(\mathcal{O}^\dagger\,\mathbb{K}\,+\,\mathbb{K}\,\mathcal{O}\ri)\psi\,.\label{var}
\ee 
This expression obviously vanishes if $\mathcal{O}$ is anti‑Hermitian, $\mathcal{O}^\dagger=-\,\mathcal{O}$, and commutes with $\mathbb{K}$. This is the case for the generators \{$H$, $P_i$, $B_i$, $J_{ij}$, $i\,n_i$, $M$\}, all of which are anti‑Hermitian and commute with the Carroll–Schr\"odinger operator \eqref{cso}. Invariance under the dilatation $D$ and the spatial special conformal transformation $K_i$ requires their Hermitian conjugates, which using \eqref{hcrules} are
\be 
D^\dagger=-\,D+2\,\omega-2\,d-1\,, \qquad\quad K_i^\dagger=-\,K_i+(2\,\omega-2\,d-1)\,x_i\,.
\ee 
Utilizing these, together with the commutation relations 
\be 
[\,D\,,\,\mathbb{K}\,]=-\,\mathbb{K}\,, \qquad \qquad\quad [\,K_i\,,\,\mathbb{K}\,]=-\,2\,x_i\,\mathbb{K}-i\,n_i\,(\omega-d+1/2)\,,
\ee
the invariance of the action \eqref{var} under the generators $D$ and $K_i$ becomes
\be 
\delta_{_{D,K_i}} S=\int dt\,d^dx~\psi^\dagger\Big(2\,\omega-2\,d+1\Big)\Big(\lambda_{_D}\,\mathbb{K} ~+~ \lambda^i_{_{K}}\,x_i\,\mathbb{K}~+~i\,\lambda^i_{_{K}}\,n_i/2\Big)\,\psi\,.\label{var2}
\ee 
This vanishes only when the dilatation weight is set to
\be
\omega=\frac{2d-1}{2}\,.\label{dw}
\ee
This can also be derived directly by demanding invariance of the action \eqref{csa} under the transformations 
\be 
t~\to~ t'=\lambda\,t\qquad  x_i~\to~ x'_i=\lambda^2\,x_i\,,\qquad
\psi~\to~ \psi'=\lambda^{-\,\omega}\,\psi\,.
\ee
Accordingly, the Carroll–Schr\"odinger action \eqref{csa} remains invariant under the transformation 
\be 
\fcolorbox{black}{gray!20}{$~~
	\begin{aligned}\phantom{\Bigg(}
		\d\psi&~=~ \bigg(~\lambda_{_H}\,H~+~\lambda^i_{_P}\,P_i~+~\lambda^i_{_B}\,B_i~+~\lambda^i_{_N}\,(i\,n_i)~+~\lambda^{ij}_{_J}\,J_{ij}~+~\lambda_{_D}\,D~+~\lambda^i_{_K}\,K_i~\bigg)\,\psi\,,\label{transf2} \phantom{\Bigg(}
	\end{aligned}
	~~$}
\ee
where $H$, $P_i$, $B_i$, $i\,n_i$, $J_{ij}$, $D$, $K_i$ are the generators of the Carroll–Schr\"odinger algebra, represented in \eqref{cscgener}, and $\lambda_{_H}$, $\lambda^i_{_P}$, $\lambda^i_{_B}$, $\lambda^i_{_N}$, $\lambda^{ij}_{_J}$, $\lambda_{_D}$, $\lambda^i_{_{K}}$ denote the corresponding transformation parameters. This completes our demonstration that the Carroll–Schr\"odinger algebra \eqref{cpcarr} is indeed the symmetry algebra of the Carroll–Schr\"odinger action \eqref{csa} in arbitrary dimensions. In particular, for $d=1$, the dilatation weight \eqref{dw} agrees with \cite{Najafizadeh:2024imn}.

\section{Two-point functions} \label{tpf}

Based on the post-Carroll algebra \eqref{pcarr} and its conformal extension \eqref{confpcarr}, this section derives a general expression for the two-point function between two complex scalar fields in a post-Carrollian CFT.

\subsection{Invariance under post-Carroll algebra generators} \label{beg}
The generators of the post-Carroll algebra are given by \eqref{ecarrg}, all of which are represented as anti-Hermitian, except for $n_i$. In Section \ref{PCT}, it was shown that the generator $n_i$ must also be represented as anti-Hermitian, specifically as $N_i = i\, n_i$ \eqref{Ni}, to maintain consistency with the field theory. As a result, this requirement implied that the fields within the post-Carrollian field theory must be complex. Therefore, following the notation used in \cite[Section 4]{Afshar:2024llh}, we consider the two-point function as 
\begin{align}
	G^{(2)}(\vec x_1,t_1;\vec x_2,t_2)=\langle0|\,\phi_{_1}(\vec x_1,t_1)~\phi^\dagger_{_2}(\vec x_2,t_2)\,|0\rangle\,.\label{two}
\end{align}
The time translation generator $H=\partial_t$ gives the differential equation
\begin{align} 
	\left(\partial_{t_1}+\partial_{t_2}\right)G^{(2)}(\vec x_1,t_1;\vec x_2,t_2)=0\,.
\end{align} 
We thus get
\begin{align} 
	G^{(2)}=G^{(2)}(\vec x_1 , \vec x_2 ; t_{12})\,,
\end{align} 
where $t_{12}=t_1-t_2$. We will address the space translation generator later, and proceed with the Carroll boost generator, $\vec B=\vec x \,\p_t$, which gives
\begin{align}  
\le(\vec x_1 \,\p_{t_1}+\vec x_2 \,\p_{t_2}\ri)G^{(2)}(\vec x_1 , \vec x_2 ; t_{12})=0\,.\label{Boostdiff}
\end{align}  
Given that $\partial_{t_1}\,G^{(2)}= \partial_{t_{12}} G^{(2)}$ and $\partial_{t_2}\,G^{(2)} =-\, \partial_{t_{12}} G^{(2)}$ for any function $G^{(2)}$, the differential equation \eqref{Boostdiff} is equivalent to 
\begin{align}  
\le(\vec x_1-\vec x_2 \ri)\p_{t_{12}}\,G^{(2)}(\vec x_1 , \vec x_2 ; t_{12})=0\,.
\end{align}  
This is solved as
\begin{align}  
G^{(2)}(\vec x_1 , \vec x_2 ; t_{12})=G(\vec x_1 , \vec x_2)+F(t_{12})\,\delta^{(d)}(\vec x_1-\vec x_2)\,,\label{twopt}
\end{align}  
where $\delta^{(d)}(\vec x):=\prod_{i=1}^d\delta(x^i)$ is the $d$-dimensional Dirac delta function. Therefore, the boost generator separates the two-point function into two distinct parts. The first term on the right-hand side of \eqref{twopt} which is only space-dependent is referred to as ``magnetic''. The second term is often called ``ultra-local'' or ``electric''.

To apply the space translation generator, let us first perform some simplifications. We express $\vec x=r\,\vec n$, where $r=|\vec x|$ and $\vec n$ is the unit radial vector. We then introduce a new variable $\widetilde G$ through
\begin{align}  
G(\vec x_1 , \vec x_2)=\frac{1}{(r_1\,r_2)^\nu}~\widetilde G(r_1,\vec n_1, r_2,\vec n_2)\,,\label{newva}
\end{align}  
where $\nu:=(d-1)/2$. Using this, together with the property of the Dirac delta function
\begin{align}  
\delta^{(d)}(\vec x_1-\vec x_2)=\frac{1}{(r_1\,r_2)^\nu}~\delta(r_1-r_2)~\delta^{(d-1)}(\vec n_1-\vec n_2)\,,\label{property}
\end{align} 
we can rewrite the two-point function \eqref{twopt} as
\begin{align} 
G^{(2)}(\vec x_1 , \vec x_2 ; t_{12})~&=~G^{(2)}(r_1,\vec n_1, r_2,\vec n_2;t_{12})\nonumber\\[9pt]
~&=~\frac{1}{(r_1\,r_2)^\nu}\le[~\widetilde G(r_1,\vec n_1, r_2,\vec n_2)~+~F(t_{12})\,\delta(r_1-r_2)\,\delta^{(d-1)}(\vec n_1-\vec n_2)~\ri]\,.\label{twop}
\end{align}  
This form will be used later. The space translation generator is defined as $\vec P=\vec\nabla$, where $\nabla_i$ is given by \eqref{nabla}. By employing the Euler operator $\vec x\cdot\vec\partial_x=r\,\partial_r$, this generator transforms into $\vec P=\vec\nabla=\vec n\le(\partial_r+\frac{\nu}{r}\ri)$. Thus, the space translation generator yields
\begin{align}
\le[~\vec n_1\le(\partial_{r_1}+\frac{\nu}{r_1}\ri)~+~\vec n_2\le(\partial_{r_2}+\frac{\nu}{r_2}\ri)\,\ri]G^{(2)}(r_1,\vec n_1, r_2,\vec n_2;t_{12})&=0\,.\label{spacetr}
\end{align}
When the form of the two-point function in \eqref{twop} is applied to this equation, the electric part becomes $F(t_{12})\,(\vec n_1 - \vec n_2)\,\partial_u \delta(u)\,\delta^{(d-1)}(\vec n_1 - \vec n_2)$, with $u = r_1 - r_2$, which automatically vanishes. Therefore, the equation \eqref{spacetr} is left with its magnetic part which is  
\begin{align}
\Big(\vec n_1\,\p_{r_1}~+~\vec n_2\,\p_{r_2}\Big)\,\widetilde G(r_1,\vec n_1, r_2,\vec n_2)=0\,.\label{magnetic}
\end{align} 
This equation has a solution when $\widetilde G$ is independent of $r_1$ and $r_2$. In addition, for $\vec n_1=\vec n_2$, the equation \eqref{magnetic} reduces to $\le(\partial_{r_1}+\partial_{r_2}\ri)\widetilde G(r_1,r_2,\vec n_1)=0$. This implies that $\widetilde G=h^-(r_1-r_2\,,\vec n_1)$. Similarly, the condition $\vec n_1=-\,\vec n_2$ yields $\widetilde G=h^+(r_1+r_2\,,\vec n_1)$. Therefore, the complete solution to the equation \eqref{magnetic} is given by
\begin{equation}\label{Carrolian2points'}
	\widetilde G(r_1,\vec n_1, r_2, \vec n_2)~=~h(\vec n_1,\vec n_2)~+~
	h^-(r_1-r_2\,,\vec n_1)\,\delta(1-\vec n_1\cdot\vec n_2)~+~
	h^+(r_1+r_2\,,\vec n_1)\,\delta(1+\vec n_1\cdot\vec n_2)
	\,,
\end{equation}
where $\vec n_1\cdot\vec n_2=\cos\theta$ with $\theta$ the angle between the two unit radial vectors, and the Dirac delta functions enforce the collinear terms $\vec n_1 =\pm \,\vec n_2$ (hence $\vec n_1\cdot\vec n_2=\pm\,1$), under which the equation reduces to the $h^{\mp}$ cases.

Let us now examine invariance under the spatial rotations $J_{ij}$. The electric part of the two-point function remains invariant under any rotation because the Dirac delta function is itself invariant (see e.g. \cite[Section 4]{Afshar:2024llh}). Therefore, we focus on the magnetic part \eqref{Carrolian2points'}. The function $h(\vec n_1,\vec n_2)$ must depend only on rotationally invariant quantities. The only such scalar formed from $\vec n_1$ and $\vec n_2$ is their dot product, $\vec n_1\cdot\vec n_2$. Thus, for rotational invariance, $h$ must be of the form $h(\vec n_1,\vec n_2)=g(\vec n_1\cdot\vec n_2)$. In addition, the functions $h^-$ and $h^+$ should not depend on $\vec n_1$, but only on $r_1-r_2$ and $r_1+r_2$ respectively. Therefore, imposing the invariance under the spatial rotations gives
\begin{equation}\label{Carrolian2points'r}
	\widetilde G(r_1,\vec n_1, r_2, \vec n_2)~=~g(\vec n_1\cdot\vec n_2)~+~
	g^-(r_1-r_2)\,\delta(1-\vec n_1\cdot\vec n_2)~+~
	g^+(r_1+r_2)\,\delta(1+\vec n_1\cdot\vec n_2)\,.
\end{equation}

\newpage
We then impose invariance under the radial direction generator $\vec N=i\,\vec n$, explained at the beginning of this section. Considering the two-point function \eqref{two}, this leads to the equation $(\vec N_{_1}+{\vec{N}_{_2}}^{\,\dagger})\,G^{(2)}=0$, or equivalently $i\!\le(\vec n_1-\vec n_2\ri)G^{(2)}=0$. Given the form of the two-point 
function in \eqref{twop}, this holds for the electric part. For the magnetic part, 
it implies that $\vec n_1 = \vec n_2$ when $\widetilde G$ is non-zero, and 
$\widetilde G = 0$ when $\vec n_1 \neq \vec n_2$. Thus, the magnetic sector \eqref{Carrolian2points'r} is constrained to take the form 
\begin{align}
\widetilde G(r_1,\vec n_1, r_2, \vec n_2)~=~\Big[~g(1)~+~g^-(r_1-r_2)~\Big]\delta(1-\vec n_1\cdot\vec n_2)\,.
\end{align}
In this way, we get the general expression for the two-point function between two complex scalar fields in a post-Carrollian field theory
\begin{align}
		G^{(2)}(r_1,\vec n_1, r_2, \vec n_2 ; t_{12})~=~\frac{1}{(r_1\,r_2)^\nu}~\bigg(~\Big[~&g(1)~+~g^-(r_1-r_2)~\Big]\delta(1-\vec n_1\cdot\vec n_2)\nonumber\\[8pt]
	&~+~F(t_{12})\,\delta(r_1-r_2)\,\delta^{(d-1)}(\vec n_1-\vec n_2)~\bigg)\,.		\label{twoppcarr}  
\end{align}
Below, we examine invariance under the generators of the conformal post-Carroll algebra \(\{D, K, K_i\}\), as represented in \eqref{gcpca}.

\subsection{K-invariance}
The invariance under the Carrollian temporal special conformal transformation generated
by $K=\vec x \cdot\vec x\,\partial_t$ is given by 
\begin{align}
	\Big(\vec x_1 \cdot\vec x_1\,\partial_{t_1}+\vec x_2 \cdot\vec x_2\,\partial_{t_2}\Big)G^{(2)}(\vec x_1 , \vec x_2 ; t_{12})=0\,,
\end{align}
or
\begin{align}
	\Big(\,r_1^2-r_2^2\,\Big)\,\partial_{t_{12}}\,G^{(2)}(r_1,\vec n_1, r_2, \vec n_2  ; t_{12})=0\,.\label{Kinvar}
\end{align}
Employing \eqref{twoppcarr}, this equation becomes
\begin{align}
	\le(\,r_1+r_2\,\ri)\le(\,r_1-r_2\,\ri)\,F\,'(t_{12})\,\delta(r_1-r_2)\,\delta^{(d-1)}(\vec n_1-\vec n_2)=0\,,
\end{align}
which is automatically satisfied. Thus, the two-point function \eqref{twoppcarr} remains invariant under $K$.

\subsection{Dilatation invariance}
Demanding invariance under the dilatation generator, $D=t\,\partial_t+\vec x\cdot\vec \partial_x+\Delta$, leads to the following differential equation
\begin{align}
\le(t_1\,\partial_{t_1}\,+\,t_2\,\partial_{t_2}
\,+\,\vec x_1\cdot\vec \partial_{x_1}\,+\,\vec x_2\cdot\vec \partial_{x_2}\,+\,\Delta_1\,+\,\Delta_2\ri)G^{(2)}(\vec x_1 , \vec x_2 ; t_{12})=0\,,
\end{align}
which can be written as 
\begin{align}
\Big(t_{12}\,\partial_{t_{12}}\,+\,r_1\,\partial_{r_1}\,+\,r_2\,\partial_{r_2}
\,+\,\Delta_1\,+\,\Delta_2\Big)\,G^{(2)}(r_1,\vec n_1, r_2, \vec n_2; t_{12})=0\,. \label{diffdilatation}
\end{align} 
To proceed, we define
\begin{align}
	\alpha=\Delta_1\,+\,\Delta_2\,-\,2\nu\,,\qquad\quad  r_{\!_{12}}=r_1-r_2\,, \label{alpha}
\end{align}
and use the property
\be 
(r_1\,\partial_{r_1}+r_2\,\partial_{r_2})\,\delta(r_1-r_2)=r_{\!_{12}}\,\partial_{r_{\!_{12}}}\,\delta(r_{\!_{12}})=-\,\delta(r_{\!_{12}})\,.
\ee
Using these, and applying \eqref{twoppcarr}, the differential equation \eqref{diffdilatation} simplifies to
\begin{align}
\Big[~\alpha\, g(1)~+~\big(r_{\!_{12}}\,\partial_{r_{\!_{12}}}+\alpha\big)\,g^-(r_{\!_{12}})~\Big]&\,\delta(1-\vec n_1\cdot\vec n_2)~=\nonumber\\[7pt]
	\Big[(1-\alpha)\,F(t_{12})\,-~t_{12}\,\partial_{t_{12}}\,F(t_{12})~\Big]&\,\delta(r_{\!_{12}})\,\delta^{(d-1)}(\vec n_1-\vec n_2)\,. \label{diffdila}
\end{align}
If $r_{\!_{12}}\neq 0$, then the right-hand side of \eqref{diffdila} vanishes, and thus for $\vec n_1=\vec n_2$ the differential equation is solved as
\be 
g^-(r_{\!_{12}})=-\,g(1)+C_1\,|r_{\!_{12}}|^{-\,\alpha}\,, \label{g-}
\ee 
where $C_1$ is an arbitrary constant. Inserting \eqref{g-} into \eqref{diffdila} shows that the left-hand side of \eqref{diffdila} vanishes everywhere except possibly at $r_{\!_{12}}=0$. Hence, the left-hand side could be at most a distribution as $\delta(r_{\!_{12}})$. Since the left-hand side is independent of $t_{12}$, evaluating \eqref{diffdila} at $\vec n_1=\vec n_2$ and $r_{\!_{12}}=0$ implies
\be 
(1-\alpha)F(t_{12})~-~t_{12}\,\partial_{t_{12}}\,F(t_{12})=0\,.
\ee 
Solving this differential equation gives
\be
F(t_{12})=C_2\,|t_{12}|^{1-\alpha}\,,\label{F}
\ee
where $C_2$ is an arbitrary factor. Therefore, substituting \eqref{g-} and \eqref{F} into \eqref{twoppcarr} yields
\begin{align}
		G^{(2)}(r_1,\vec n_1, r_2, \vec n_2 ; t_{12})~=~\frac{1}{(r_1\,r_2)^\nu}\,\bigg[~
	&~C_1\,|r_{\!_{12}}|^{-\,(\Delta_1+\Delta_2-d+1)}~\delta(1-\vec n_1\cdot\vec n_2)~+\nonumber\\[8pt]
	&~C_2\,|t_{12}|^{-\,(\Delta_1+\Delta_2-d)}~\delta(r_{\!_{12}})\,\delta^{(d-1)}(\vec n_1-\vec n_2)~\bigg]\,.\label{D-invar}
\end{align}
This is the general expression of the two-point function for two complex scalar fields of scaling dimensions $\Delta_1$ and $\Delta_2$. In the terminology of \cite{Afshar:2024llh}, this form is invariant under the conformal post-Carroll algebra of type D-K. However, it is not yet invariant under $K_i$.

\subsection{K$_i$ invariance}
Invariance under the spatial special conformal transformation, $\vec K=2\,\vec x\,(t\,\partial_t+\vec x\cdot\vec\partial_x+\Delta)\,-\,x^2\,\vec\nabla$, with $\nabla_i$ given in \eqref{nabla}, yields the following differential equation
\begin{align}
	\Big[~&\vec x_1\,(t_1\,\partial_{t_1}+\vec x_1\cdot\vec\partial_{x_1}+\Delta_1)\,-\,\frac{1}{2}\,(\vec x_1\cdot\vec x_1)\,\vec\nabla_{1}~+\nonumber\\[8pt]
	&\vec x_2\,(t_2\,\partial_{t_2}+\vec x_2\cdot\vec\partial_{x_2}+\Delta_2)\,-\,\frac{1}{2}\,(\vec x_2\cdot\vec x_2)\,\vec\nabla_{2}~\Big]\,G^{(2)}(\vec x_1 , \vec x_2 ; t_{12})=0\,.\label{K-in}
\end{align}
For $r_{\!_{12}}\neq 0$, the second line of \eqref{D-invar} vanishes; hence, the remaining term (i.e. the magnetic part) can be substituted into \eqref{K-in}. Applying this substitution, and using $\vec\nabla=\vec n\le(\partial_r+\frac{\nu}{r}\ri)$, the equation \eqref{K-in} reduces to 
\begin{align}
	\le[~\vec n_1\,r_1\,\le(\,\frac{1}{2}\,r_1\,\partial_{r_1}-\frac{\nu}{2}+\Delta_1\ri)+\,\vec n_2\,r_2\,\le(\,\frac{1}{2}\,r_2\,\partial_{r_2}-\frac{\nu}{2}+\Delta_2\ri)~\ri]\,\frac{\,C_1\,|r_{\!_{12}}|^{-\,\alpha}}{(r_1\,r_2)^\nu}~\delta(1-\vec n_1\cdot\vec n_2)=0\,,\label{kin}
\end{align}
where $\vec x=r\,\vec n$ and the Euler operator $\vec x\cdot\vec\partial_x= r\,\partial_r$ have been used. Since the Dirac delta function enforces $\vec n_1=\vec n_2$, the equation \eqref{kin} reduces to 
\be 
C_1\le[\,\Delta_1\,r_1\,+\,\Delta_2\,r_2\,-\,\le(\nu+\frac{\alpha}{2}\ri)(r_1+r_2)\,\ri]=0\,,
\ee 
which, using \eqref{alpha}, is satisfied iff either $\Delta_1=\Delta_2\equiv\Delta$ or $C_1=0$. Therefore, the ``magnetic'' part of the two-point function for distinct scaling dimensions is zero.

For $r_{\!_{12}}=0$, the second line of \eqref{D-invar} (i.e. the electric part) is used in \eqref{K-in}. In this case, since the Dirac delta function again enforces $\vec n_1=\vec n_2$, the equation \eqref{K-in} reduces to
\begin{align}
	C_2\le[ \big(r_1\,t_1-r_2\,t_2\big)\partial_{t_{12}}+\frac{1}{2}\,r_1^2\,\partial_{r_1}+\frac{1}{2}\,r_2^2\,\partial_{r_2}+r_1\,\Delta_1+r_2\,\Delta_2-\frac{\,\nu\,}{2}\,(r_1+r_2) \ri]\frac{~|t_{12}|^{1-\alpha}}{(r_1\,r_2)^\nu}~\delta(r_{\!_{12}})=0.
\end{align}
Introducing $\tilde{r}_{\!_{12}}=r_1+r_2$, and using $r_{\!_{12}}\,\delta(r_{\!_{12}})=0$, this equation simplifies to
\begin{align}
		C_2\le[~\frac{1}{2}\,\tilde{r}_{\!_{12}}\Big(t_{12}\,\partial_{t_{12}}+\Delta_1+\Delta_2-\nu\Big)
		+\frac{1}{2}\,\tilde{r}_{\!_{12}}\,r_{\!_{12}}\,\partial_{r_{\!_{12}}}~\ri]|t_{12}|^{1-\alpha}\,\delta(r_{\!_{12}})=0. 
\end{align}
Since $r_{\!_{12}}\,\partial_{r_{\!_{12}}}\,\delta(r_{\!_{12}})=-\,\delta(r_{\!_{12}})$, this further reduces to
\begin{align}
	C_2\,\Big(\Delta_1+\Delta_2-\nu-\alpha\Big)\,\delta(r_{\!_{12}})=0\,,
\end{align}
which, using \eqref{alpha}, holds iff either $d=1$ or $C_2=0$. Thus, the ``electric'' part of the two-point function in $d>1$ is zero. However, in $d=1$, it is non-zero for scalar fields of either equal or distinct scaling dimensions, since the invariance under $K_i$ does not constrain the scaling dimensions in the electric sector. 

Therefore, we find that in a post-Carrollian CFT in $1+1$ spacetime dimensions invariant under the conformal post-Carroll algebra \eqref{confpcarr}, the two-point function between two complex (or real\,\footnote{As noted in Section \ref{PCT}, the generator $n_i$ is trivially realised in $1+1$ dimensions, allowing the fields to be either real or complex. Hence, one may consider \eqref{two} with real scalar fields, and the result \eqref{d=1} holds for both cases.}) scalar fields with the same scaling dimension $\Delta$ and that between two scalar fields with distinct scaling dimensions $\Delta_1\neq\Delta_2$ are given respectively by
\be
\text{For}~d=1\quad\fcolorbox{black}{gray!20}{$~~~
	\begin{aligned}
	\phantom{\Bigg(}
	G^{(2)}(r_{\!_{12}}, t_{_{12}}) ~=~ 
	\begin{cases}
		\displaystyle   ~\frac{C_1}{~|r_{\!_{12}}|^{\,2\Delta}~}~+~\frac{C_2}{~|t_{_{12}}|^{\,2\Delta\,-\,1}~}~\delta(r_{\!_{12}})\,, & \qquad \Delta_1=\Delta_2\equiv\Delta\,, \\[20pt]
		\displaystyle ~\frac{C_2}{~|t_{_{12}}|^{\,\Delta_1\,+\,\Delta_2\,-\,1}~}~\delta(r_{\!_{12}})\,, & \qquad \Delta_1 \neq \Delta_2\,.
	\end{cases}
	\phantom{\Bigg(}
	\end{aligned}\label{d=1}
	~~~$}
\ee
This agrees with \cite{Afshar:2024llh} in $1+1$ dimensions, as expected. As noted earlier, the Carroll algebra and the post-Carroll algebra share the same algebraic structure in $1+1$ dimensions, so their correlation functions must be equivalent. 

In higher dimensions $d>1$, however, we find that the electric part vanishes, and the magnetic part with distinct scaling dimensions $\Delta_1 \neq \Delta_2$ also vanishes, while the magnetic part with the same scaling dimension $\Delta_1=\Delta_2\equiv\Delta$ survives. Therefore, in a post-Carrollian CFT in $1+d$ spacetime dimensions (with $d>1$) invariant under the conformal post-Carroll algebra \eqref{confpcarr}, the two-point function between two complex scalar fields with scaling dimensions $\Delta_1$ and $\Delta_2$ becomes
\be 
\hspace{-.2cm}\text{For}~d>1\quad\fcolorbox{black}{gray!20}{$~~
	\begin{aligned}
		\phantom{\Bigg(}
		\!\!\!\!	G^{(2)}(r_1, r_2, \vec n_1\cdot\vec n_2)~=~\delta_{{\Delta_1,\Delta_2}}~\frac{C_1}{\le(r_1\,r_2\ri)^{\,\frac{d-1}{2}}}~\frac{1}{~|r_1-r_2|^{\,\Delta_1\,+\,\Delta_2\,-\,d\,+\,1}}~\delta(1-\vec n_1\cdot\vec n_2)\,.\label{d>1}
		\phantom{\Bigg(}
	\end{aligned}
	~$} 
\ee
Therefore, in higher dimensions ($d>1$), only the magnetic sector of the two-point function survives. 

\section{Conclusions and outlook} \label{concl}

In this work, we extended the Carrollian framework by incorporating leading \(c\)-dependent corrections to the Carroll transformations, which yielded the extended Carroll transformations \eqref{epmt}. Then, by employing post-Carrollian mechanics, these transformations led us to the introduction of post-Carroll transformations \eqref{epmtp} for energy and momentum. We demonstrated that the post-Carroll transformations are fully consistent with post-Carrollian mechanics \cite{Najafizadeh:2025ksm}, such that relations within this framework transform covariantly under these transformations, see e.g. \eqref{cova}.

Using the post-Carroll transformations, we derived two novel algebraic structures in arbitrary spacetime dimensions: $(i)$ The post-Carroll algebra \eqref{pcarr}, \(\mathfrak{pcarr}(d+1)\), which generalizes the Carroll algebra through the inclusion of a radial direction generator \(n_i\) and a modified representation of the space translation generator \(P_i = \nabla_i\). $(ii)$ The Carroll–Bargmann algebra \eqref{pcarrb}, \(\mathfrak{carrb}(d+1)\), a central extension of the post-Carroll algebra by a central charge \(M\). This algebra overcomes the limitation of the standard Carroll algebra, namely the absence of a nontrivial central charge in higher dimensions.

We further constructed conformal extensions of both algebras. The conformal post-Carroll algebra \eqref{confpcarr}, \(\mathfrak{cpcar}(d+1)\), with critical exponent \(z=1\), was obtained by including the special conformal transformation generators \(D\), \(K\) and \(K_i\). More significantly, we built the conformal extension of the Carroll–Bargmann algebra with conformal generators $D$ and $K_i$, resulting in the Carroll–Schr\"odinger algebra \eqref{cpcarr}, \(\mathfrak{carrsch}(d+1)\), with critical exponent \(z=1/2\). We demonstrated that this algebra precisely matches the symmetry algebra of the higher-dimensional Carroll–Schr\"odinger field theory \cite{Najafizadeh:2024imn}, a symmetry that was absent in the literature so far, although it was well established in \(1+1\) dimensions.

On the field-theoretic side, we investigated actions invariant under these algebras. These algebras were shown to require complex scalar fields due to the presence of the radial generator \(n_i\) (or more precisely the anti-Hermitian generator \(N_i = i n_i\)), except in \(1+1\) dimensions where real fields become admissible. Accordingly, we constructed the electric \eqref{epca} and magnetic \eqref{mpca} post-Carroll actions and studied their invariance under the conformal post-Carroll algebra.

Finally, based on the conformal post-Carroll algebra \eqref{confpcarr}, we derived the general form of two-point functions between two complex scalar fields in a post-Carrollian CFT. Our analysis revealed: in \(1+1\) dimensions, both electric and magnetic sectors contribute, with the electric sector surviving even for distinct scaling dimensions \eqref{d=1}. In higher dimensions (\(d>1\)), however, the electric sector vanishes entirely, and the magnetic sector survives only when the two complex scalar fields share the same scaling dimension \eqref{d>1}.

Several avenues remain open for future investigation in the post-Carrollian regime, analogous to the studies already performed in the Carrollian regime. For instance, it would be interesting to investigate higher-dimensional extensions of the conformal post-Carroll algebra \eqref{confpcarr} in 1+2 spacetime dimensions and beyond. We note that, in 1+1 dimensions, its higher-dimensional extension is equivalent to the Carrollian conformal algebra (CCA), as they share the same structure in this dimension. Moreover, it would be interesting to study the construction of a full post-Carrollian gravity in curved backgrounds and find a possible connection with post-Carrollian gravity in flat spacetime \cite{Najafizadeh:2025ksm}. Besides, it would be interesting to apply the method of \cite{Afshar:2015aku} to the Carroll–Schr\"odinger algebra \eqref{cpcarr} and to determine whether a \(z=1/2\) Ho\v{r}ava–Lifshitz gravity can exist. In addition, exploring post-Carroll fermions and investigating supersymmetric extensions — both at the level of algebras and field theory — would be worthwhile. Furthermore, the classification of all possible conformal extensions of the post-Carroll algebra, along the lines of \cite{Afshar:2024llh}, could reveal additional critical exponents and exotic algebras. Finally, the role of post-Carrollian symmetry in flat-space holography and celestial holography remains an open and promising direction.

\paragraph{Acknowledgments:} We are grateful to Shahin Sheikh-Jabbari for his support and encouragement. We also thank Mohammad Khorrami, Daniel Grumiller, and Peter Horvathy for helpful discussions, comments, and correspondence. 

\appendix

\section{Useful relations} \label{useful}
Let us take into account the following relations
\begin{align}
&n_i=\frac{x_i}{r}\,, &&\nabla_i=\frac{n_i}{\,r\,}\le(\vec x\cdot\vec\partial_x~+~\frac{d-1}{2}\ri)\,,&&\nabla_x=n^{i}\,\nabla_i\,,
\end{align}
defined in \eqref{N}, \eqref{nabla} and \eqref{nablax}. Using these, we obtain the following zero commutation relations
\begin{align}
 &[\,\nabla_i\,,\,\nabla_j\,]=0\,, && [\,\nabla_i\,,\,n_j\,]=0\,, \\[10pt]
 &[\,\nabla_x\,,\,\nabla_i\,]=0\,, && [\,\nabla_x\,,\,n_i\,]=0\,, \\[10pt]
 &[\,\nabla_x\,,\,\nabla_x\,]=0\,, && [\,n_i\,,\,x\cdot\p\,]=0\,, \\[10pt]
 &[\,\nabla_x\,,\,J_{ij}\,]=0\,, &&[\,\nabla_i\nabla^{\,i}\,,\,J_{jk}\,]=0\,.
\end{align}
In addition, we find the useful non-zero commutation relations as follows:
\begin{align}
	&[\,\nabla_i\,,\,x_j\,]=n_i\,n_j\,, &&[\,\nabla_i\,,\,\p_j\,]=\frac{1}{r}\,\big(\,2\,n_i\nabla_j-n_i\,\p_j-\delta_{ij}\,\vec n\cdot\vec\nabla\,\big)\,, \\[10pt]
	&[\,\nabla_i\,,\,\vec x\cdot\vec\partial_x\,]=\nabla_i\,, &&[\,\p_i\,,\,n_j\,]=\frac{1}{r}\,\big(\delta_{ij}-n_i\, n_j\big)\,,\\[10pt]
    &[\,\nabla_i\,,\,x^2\,]=2\,x_i\,,&&[\,\nabla_i\,,\,J_{jk}\,]=\frac{\!}{\!}\d_{ij}\nabla_{k}-\d_{ik}\nabla_{j}\,,\\[10pt]
	&[\,\nabla_x\,,\,x_i\,]=n_i\,,&&[\,n_i\,,\,J_{jk}\,]=\frac{\!}{\!}\d_{ij}\,n_{k}-\d_{ik}\,n_{j}\,,\\[10pt]
	&[\,\nabla_i\,,\,r\,]=n_i\,,&&[\,\nabla_x\,,\,\p_i\,]=\frac{1}{r}\,\le(\nabla_i-\p_i\ri)\,,\\[10pt]
	&[\,\p_i\,,\,r\,]=n_i\,, && [\,\vec x\cdot\vec\partial_x\,,\,\frac{n_i}{r}\,]=-\,\frac{n_i}{r}\,,\\[10pt]
	&[\,\p_i\,,\,\frac{1}{r}\,]=-\,\frac{n_i}{\,r^{\,2}\,}\,, && [\,\p_i\,,\,\frac{n_j}{r}\,]=\frac{1}{r^{\,2}}\,\big(\delta_{ij}-2\,n_i \,n_j\big)\,.
\end{align}
Furthermore, we find
\begin{align}
{\nabla_i\nabla^{\,i}}
	~=~\vec\nabla\cdot\vec\nabla~=~\vec\nabla^2~=~\frac{1}{r^2}\le[(\vec x\cdot\vec\partial_x)^2~+~(d-2)(\vec x\cdot\vec\partial_x)~+~\frac{(d-1)(d-3)}{4}\ri]\,.\label{nabnab}
\end{align}
Using the Euler operator $\vec x\cdot\vec\partial_x=r\,\partial_r$, this is equivalent to 
\begin{align}
	\vec\nabla^2~=~\frac{1}{r^{\,d-1}}~\frac{\p}{\p r}\le(r^{\,d-1}\,\frac{\p}{\p r}\ri)~+~\frac{(d-1)(d-3)}{4\,r^2}\,,\label{nabnab2}
\end{align}
which in $d=3$ gives the radial part of the standard Laplacian $\vec{\bm{\nabla}}^2$.

\section{Galilei transformations and beyond} \label{gt}

In this section, we first briefly review the derivation of the Galilei transformations, and then go beyond them by reviewing the Newton transformations. Let us consider the Lorentz transformations for energy and momentum, given in \eqref{ltem}. In the strict limit $c\to \infty$, where $\g\to 1$, these transformations reduce to the ``Galilei transformations''
\begin{subequations}\label{gemt}
	\begin{align}
		E^{\,'}&=E\,-\,\vec{u}\cdot\vec{p}\,, \label{get}\\[3pt]
		\vec{p}^{~'}&=\vec{p}\,,\label{gmt}
	\end{align}	
\end{subequations}
where here $\vec{u}$ is the Galilei boost parameter. The transformation rule \eqref{get} implies that energy does change under a Galilei boost, as expressed by the commutator \([H, U_i] = P_i\), where $H$ is the Hamiltonian, $U_i$ the Galilei boost generator, and $P_i$ the space translation generator. In addition, the transformation \eqref{gmt} demonstrates that momentum remains invariant under a Galilei boost, yielding \([P_i, U_j] = 0\). When rotation generators are included, the space translation and the boost transform as vectors under spatial rotations \( J_{ij} \). Therefore, these generators \{\( H \), \( P_i \), \( U_i \), \( J_{ij} \)\}, represented as \be 
H=\p_t\,,\qquad\qquad P_i=\p_i \,, \qquad\qquad  U_i=t\,\p_i\,,\qquad\qquad  J_{ij} = x_i\,\p_j-x_j\,\p_i \,,
\ee 
define the ``Galilei algebra'' with the non-zero commutation relations:
\begin{align}
	[\,H\,,\,U_i\,]=P_i\,,\qquad[\,P_i\,,\,J_{jk}\,]=\d_{i[j}\,P_{k]}\,,\qquad[\,U_i\,,\,J_{jk}\,]=\d_{i[j}\,U_{k]}\,,\qquad[\,J_{ij}\,,\,J_{kl}\,]=\d_{[i[k}J_{l]j]}\,.\label{Galileia}
\end{align}
Here, the space translation generator and the boost commute $[\,P_i\,,\, U_j\,] = 0$. However, as one can show, it is possible to introduce a non-vanishing commutator $[\,P_i\,,\, U_j\,] \ne 0$. The price to pay is the inclusion of $c$-dependent corrections in the Galilei transformations \eqref{gmt}, leading to a central extension of the algebra, i.e. to the Bargmann algebra. The presence of such corrections indicates that one is dealing with Newtonian particles, for which Newtonian mechanics must be employed. Therefore, before including corrections to the Galilei transformations, let us first list quantities in Newtonian mechanics in Subsec. \ref{pcmeN} and then reproduce the Bargmann algebra and the Newton transformations in Subsec. \ref{barg}.

\subsection{Newtonian mechanics} \label{pcmeN}

In this framework, the total energy is given by 
\begin{align}
	E&\,=\,E_{_0}~+~E_{_N}\,,\label{total mom}
\end{align}
where 
\be 
E_{_0}=m\,c^2\,,\label{e0}
\ee 
is the rest energy. The Newtonian kinetic energy $E_{_N}$ and the Newtonian momentum $\vec{p}_{_N}$ are given by
\begin{align}
	E_{_N}&=\frac{1}{2}\,m\,v^2\,,\label{en}\\[5pt]
	\vec{p}_{_N}&=m\,\vec{v}\,, \label{pn}
\end{align}
and thus the energy-momentum relation reads
\be 
E_{_{N}}\,=\,\frac{(\,	{\vec{p}}_{_N}\,)^2}{2m}\,.\label{emrN}
\ee 
As we will see in the main text, the quantities \eqref{total mom}–\eqref{emrN} in the Newtonian framework have counterparts \eqref{tp}–\eqref{pcem} in the post-Carrollian framework.

\subsection{Newton transformations} \label{barg}

To derive the Bargmann algebra and the Newton transformations, we include the leading $c$-dependent corrections to the Galilei transformations \eqref{gemt}. To this end, we consider again \eqref{ltem} and expand $\gamma$ in powers of $(u/c)^2$, that is $\g=1+u^2/(2\,c^2)+\mathcal(c^{-\,4})$. The resulting transformations, which may be referred to as ``expanded Galilei transformations'', are given by 
\begin{subequations}\label{gepmt}
	\begin{align}
		E^{\,'}&=E\,-\,\vec{u}\cdot\vec{p}\,+\,\frac{u^2}{2}\,\frac{E}{c^2}\,, \label{1gepmt}\\[5pt]
		\vec{p}_{\,\shortparallel}^{~'}&=\vec{p}_{\,\shortparallel}\,-\,\vec{u}~\frac{E}{c^2}\,, \label{2gepmt}\\[5pt] 
		{\vec{p}_{\!_\perp}}^{~'}&=\vec{p}_{\!_\perp}\,. \label{3gepmt}
	\end{align}
\end{subequations}
As expected, these reduce to the Galilei transformations \eqref{gemt} in the strict limit $c\to\infty$. Here, the $c$-dependent corrections reveal an important observation: unlike the Galilei case \eqref{gmt}, where momentum remains invariant under a boost, the transformation in \eqref{2gepmt} illustrates that momentum is not invariant under a Galilei boost when aligned parallel to the boost direction. This happens once $c$-dependent corrections are taken into account. To see the advantage, we substitute the total energy \eqref{total mom} into \eqref{2gepmt}, from which the leading term yields
\be 
\vec{p}_{\,\shortparallel}^{~'}=\vec{p}_{\,\shortparallel}\,-\,\vec{u}~m\,.\label{gc}
\ee 
This demonstrates that a central charge $M$ must be included on the right-hand side of the commutator between space translation and boost 
\be 
[\,P_i\,,\,U_j\,]=\delta_{ij}\,M\,.\label{M}
\ee 
As a result, upon including the central charge $M$ (which is equivalent to considering the leading $c$-dependent corrections), the Galilei generators extend to the Bargmann generators $\{ H, P_i, U_i, J_{ij}, M \}$, represented by
\be
H=\p_t\,,\quad\qquad P_i=\p_i \,, \quad\qquad  U_i=t\,\p_i+x_i\,M\,,\quad\qquad  J_{ij} = x_i\,\p_j-x_j\,\p_i\,, \quad\qquad M=-\,im\,. \label{BargmGe}
\ee 
Here, we represent the central charge in such a way that $M$, the boost generator $U_i$, and indeed all generators are anti-Hermitian. These generators give rise to the ``Bargmann algebra'', with the non-zero commutation relations:
\begin{align}
	&[\,H\,,\,U_i\,]=P_i\,, &&[\,P_i\,,\,J_{jk}\,]=\d_{i[j}\,P_{k]}\,, &&[\,J_{ij}\,,\,J_{kl}\,]=\d_{[i[k}J_{l]j]}\,, \nonumber\\[5pt] 
	&[\,P_i\,,\,U_j\,]=\d_{ij}\,M\,, &&[\,U_i\,,\,J_{jk}\,]=\d_{i[j}\,U_{k]}\,. && \label{Barga} 
\end{align}

Let us now proceed to derive the Newton transformations from \eqref{gepmt}. The resulting transformations satisfy the requirements of Newtonian mechanics. We substitute the total energy \eqref{total mom} into the $c$-dependent transformations, \eqref{1gepmt} and \eqref{2gepmt}, and retain only the leading order terms. Since $\vec u\cdot\vec{p}_{\!_\perp}=0$, only the parallel component contributes; thus $\vec u\cdot\vec p=\vec u\cdot\vec{p}_{\,\shortparallel}=\vec u\cdot{\vec{p}}_{_N}$, where we identified $\vec{p}_{\,\shortparallel}={\vec{p}}_{_N}$. In addition, noting that the rest energy is boost-invariant, i.e. $E_{_0}^{\,'}=E_{_0}=m\,c^2$, this term cancels from both sides of \eqref{1gepmt}. Accordingly, one obtains the ``Newton transformations'' 
\begin{subequations}\label{gs1}
	\begin{align} 
		E_{_{N}}^{\,'}&\,=\,E_{_{N}} \,-\, \vec u\cdot{\vec{p}}_{_N} \,+\, \frac{1}{2}\,m\,{u}^2\,,\\[5pt]
		{\vec{p}}_{_N}^{~'}&\,=\,{\vec{p}}_{_N} \,-\, m\,\vec{u}\,.
	\end{align} 
\end{subequations}
Using \eqref{pn}, the latter directly implies the velocity transformation
\be 
\vec{v}^{~'}=~\vec{v}\,-\,\vec{u}\,. \label{v}
\ee
As a result, under the Newton transformations \eqref{gs1}, the energy-momentum relation \eqref{emrN} transforms covariantly; that is
\be 
E_{_{N}}^{\,'}=\frac{(\,{\vec{p}}_{_N}^{~'}\,)^2}{2m}\qquad \longleftrightarrow\qquad E_{_{N}}=\frac{(\,	{\vec{p}}_{_N}\,)^2}{2m}\,.
\ee 
Using \eqref{en} and \eqref{pn}, the transformations in \eqref{gs1} can be expressed equivalently as
\begin{subequations}\label{gs}
	\begin{align} 
		E_{_{N}}^{\,'}&=\frac{1}{2}\,m\,(\vec{v}-\vec{u}\,)^2\,,\\[5pt]
		{\vec{p}}_{_N}^{~'}&=m\,(\vec{v}-\vec{u}\,)\,.
	\end{align} 
\end{subequations}
Together with the velocity transformation \eqref{v}, this shows that the Newtonian kinetic energy \eqref{en} and momentum \eqref{pn} transform covariantly as well.

\section{Derivation of post-Carroll transformations} \label{Apct}

This appendix details the derivation of the post-Carroll transformations \eqref{epmtp}. To derive the energy transformation \eqref{1epmtp}, we begin with \eqref{1epmt}. Generally, the energy $E=E_{\c}+E_{\pc}$ can be decomposed into the energy of a magnetic Carroll particle $E_{\c}$ and that of a post-Carroll particle $E_{\pc}$. However, since magnetic Carroll particles have zero energy $E_{\c}=0$, it follows that $E = E_{\pc}$. Thus, the relation \eqref{1epmt} reduces to
\be 
E^{\,'}_{\pc} = E_{\pc} - c^2\,\vec{b}\cdot(\vec{p}_{\c}+\vec{p}_{\pc})\,,\label{etra}
\ee 
where the total momentum \eqref{tp} is substituted. Plugging the momentum of a magnetic Carroll particle $\vec{p}_{\c}$ \eqref{cm} and that of a post-Carroll particle $\vec{p}_{\pc}$ \eqref{pcm} into the latter, and keeping only the leading term (ignoring the $\mathcal{O}(c^5)$ contribution), one finds 
\be 
E^{\,'}_{\pc} = E_{\pc} \,-\, \vec{b}\cdot\vec{v}~\Big(\,\frac{mc^3}{v}\,\Big)\,.
\ee 
Using \eqref{pcen}, this expression becomes the energy transformation \eqref{1epmtp} in the post-Carrollian framework.

To derive the momentum transformation \eqref{2epmtp}, we begin with \eqref{compact} and substitute the total momentum \eqref{tp}, which yields
\be
\vec{p}^{~'}_{\pc}\,=\,\vec{p}_{\pc}\,-\,\vec{b}\,E\,+\,\tfrac{1}{2}\,c^2\big(\,\vec{b}\cdot\vec{p}_{\c}\,+\,\vec{b}\cdot\vec{p}_{\pc}\,\big)\,\vec{b}\,.\label{compact1}
\ee
Here, we assumed the fact that the momentum of a magnetic Carroll particle is boost-invariant. This is evident from \eqref{mt}, since magnetic Carroll particles carry zero energy; hence, $\vec{p}_{\c}{}^{'}=\vec{p}_{\c}=mc\,\hat{v}$. This invariance allowed us to eliminate the magnetic Carroll contribution from both sides of \eqref{compact1}, leaving only the post-Carroll transformation. Using \eqref{cm}, \eqref{pcen}, and \eqref{pcm} which shows $\vec{p}_{\pc}=|\vec{p}_{\pc}|\,\hat{v}$, the relation \eqref{compact1} up to the leading term (neglecting the $\mathcal{O}(c^5)$ term) gives
\begin{align}
\vec{p}^{~'}_{\pc}&=~\vec{p}_{\pc}\,-\,\vec{b}~\Big(\frac{mc^3}{v}\Big)\,+\,\tfrac{1}{2}\,c^2\big(\,\vec{b}\cdot\hat{v}\,mc\,\big)\,\vec{b}\\[8pt]
&=~\vec{p}_{\pc}\,-\,2\,\vec{b}\,v~\Big(\frac{mc^3}{\,2v^{\,2}\,}\Big)\,+\,\Big(\frac{mc^3}{\,2v^{\,2}\,}\Big)\,v^2\big(\,\vec{b}\cdot\hat{v}\,\big)\,\vec{b}\\[8pt]
&=~\vec{p}_{\pc}\,-\,2\,\vec{b}\,v~|\vec{p}_{\pc}|\,+\,\vec{b}\,v~|\vec{p}_{\pc}|\,\big(\,\vec{b}\cdot\vec{v}\,\big)\,.\label{222}
\end{align} 
Since $\vec{b}\cdot\vec{v}=b\,v\,\cos\alpha$, where $b=|\vec{b}|$ and $\alpha$ is the angle between the boost $\vec{b}$ and the velocity $\vec{v}$, one has $\vec{b}\cdot\hat{v}=b\,\cos\alpha$, or equivalently $\vec{b}=\hat{v}\,b\cos\alpha$. Using this, the relation \eqref{222} simplifies to
\begin{align}
\vec{p}^{~'}_{\pc}&=~\vec{p}_{\pc}\le(\,1\,-\,2\,b\,v\,\cos{\alpha}\,+\,b^2\,v^2\,\cos^2{\!\alpha}\,\ri)\\[8pt]
&=~\vec{p}_{\pc}\le(\,1\,-\,b\,v\,\cos{\alpha}\,\ri)^2=\vec{p}_{_\mathrm{pc}}\,\big(\,1-\vec{b}\cdot\vec{v}\,\big)^2\,,\label{ee}
\end{align}
which is the momentum transformation \eqref{2epmtp} in the post-Carrollian regime. 

For our subsequent purpose applied in Section \ref{beyond}, let us consider \eqref{ee} up to the first order of the boost parameter, which is
\begin{align}
	\vec{p}_{\pc}^{~'} &\,=\, \vec{p}_{\pc}\big(1\,-\,2\,\vec{b}\cdot\vec{v}\,\big)\,.\label{lo}
\end{align}
Using \eqref{pcen} and \eqref{pcm}, this becomes
\begin{align}
	\vec{p}_{\pc}^{~'} 	&\,=\,\vec{p}_{\pc}\,-\,\Big(\frac{mc^3}{2v^{\,2}}\Big)\,\hat{v}\,(\,2\,\vec{b}\cdot\vec{v}\,)\\[8pt]
	&\,=\,\vec{p}_{\pc}\,-\,\Big(\frac{mc^3}{v}\Big)\,(\vec{b}\cdot\hat{v})\,\hat{v}\\[8pt]
	&\,=\,\vec{p}_{\pc}\,-\,E_{\pc}\,(\vec{b}\cdot\hat{v})\,\hat{v}\,.	\label{1order}
\end{align}

\bibliographystyle{hephys}
\small\bibliography{references}

\end{document}